\documentclass[11pt, a4paper]{book}
\usepackage{algorithm}
\usepackage[noend]{algpseudocode}
\usepackage{mathptmx}
\usepackage{graphics}

\usepackage[palatino]{anuthesis}

\usepackage[final]{prelim2e}  

\usepackage[square]{natbib}

\usepackage[T1]{fontenc}
\usepackage[scaled]{PTSans}          
\usepackage[scaled]{DejaVuSansMono}  

\usepackage[stretch=10,shrink=10]{microtype}

\usepackage{amsmath}
\usepackage{amssymb}

\usepackage{graphicx}
\usepackage[table]{xcolor}
\usepackage{float}

\usepackage{caption}
\usepackage{subcaption}
\usepackage{booktabs}
\usepackage{relsize}
\usepackage{multirow}
\usepackage{array}

\usepackage{listings}

\usepackage{enumitem}

\definecolor{todocolor}{rgb}{0.8,0.8,1.0}
\usepackage[color=todocolor,colorinlistoftodos]{todonotes}

\usepackage[binary-units=true]{siunitx}

\usepackage{xspace}
\usepackage{afterpage}
\usepackage[normalem]{ulem}
\usepackage{makeidx}
\usepackage{url}
\usepackage{pdflscape}
\PackageWarning{report template}{Update your PDF title and author!} 
\usepackage[
  hyperindex=true,
  bookmarks=true,
  pdftitle={YOUR TITLE HERE},
  pdfauthor={YOUR NAME HERE},
  colorlinks=false,
  plainpages=false,
  pdfborder={0 0 0},
  pdfpagelabels,
  hyperfootnotes=false]{hyperref}
\usepackage[all]{hypcap}  

\usepackage[hyperpageref]{backref}
\renewcommand*{\backref}[1]{}
\renewcommand*{\backrefalt}[4]{
  \ifcase #1 %
  \or
    (cited on page #2)%
  \else
    (cited on pages #2)%
  \fi
}

\usepackage[nameinlink,capitalise]{cleveref}

\newcommand{\doi}[1]{\href{http://dx.doi.org/#1}{\nolinkurl{doi:#1}}}

\lstdefinestyle{default}{
  numbers=left,
  numberstyle=\tiny,
  stepnumber=1,
  numbersep=2em,
  language=java,                         
  basicstyle=\footnotesize\ttfamily,     
  commentstyle=\itshape,                 
  stringstyle=\ttfamily,
}
\crefname{line}{line}{lines}  

\definecolor{tableheadcolor}{rgb}{0.8,0.8,1.0}
\definecolor{tablealtcolor}{rgb}{0.9,0.9,0.95}

\newcolumntype{B}{>{\global\let\currentrowstyle\relax}}
\newcolumntype{^}{>{\currentrowstyle}}
\newcommand{\rowstyle}[1]{\gdef\currentrowstyle{#1}#1\ignorespaces}
\newcolumntype{C}{>{\bfseries}}

\let\oldtabular\tabular
\let\endoldtabular\endtabular
\renewenvironment{tabular}{\sffamily\oldtabular}{\endoldtabular}

\makeatletter
\renewcommand\bibsection{%
      \chapter*{\bibname\@mkboth{\bibname}{\bibname}}}
\makeatother

\AtBeginDocument{
  \DeclareSymbolFont{AMSb}{U}{msb}{m}{n}
  \DeclareSymbolFontAlphabet{\mathbb}{AMSb}}

\newcommand{\icode}[1]{{\lstset{basicstyle=\ttfamily\small}\lstinline@#1@}}

\definecolor{fixcolor}{rgb}{1,0.8,0.8}

\definecolor{commentcolor}{rgb}{0.8,1.0,0.8}

\title{Title Is Here}
\author{Author Name}
\date{\today}

\renewcommand{\thepage}{\roman{page}}

\makeindex
\begin{document}

\pagestyle{empty}
\thispagestyle{empty}
\input{titlepage}

\vspace*{7cm}
\begin{center}
  Except where otherwise indicated, this thesis is my own original
  work.
\end{center}

\vspace*{4cm}

\hspace{8cm}\makeatletter Yixin Cheng\makeatother\par
\hspace{8cm} 19 November 2021

\cleardoublepage
\pagestyle{empty}
\chapter*{Acknowledgments}
\addcontentsline{toc}{chapter}{Acknowledgments}
Firstly, I must express my sincere appreciation to my supervisor, Dr. Bernardo Pereira Nunes, for his heuristic guidance, perceptive comments, and personal support. It is an honor as well as a luck to be his student. His enthusiasm and meticulousness towards academics motivate me along the way. He has monitored my progress and offered advice and encouragement throughout. Without his constant help, I would not have completed this project on schedule.
\\
\\
Thanks also to Felipe Cueto-Ramirez, for providing support and advice for my project.
\\
\\
Special thanks to Jinxiao Xie, who has always encouraged me to challenge myself and made time to help and support me.
\\
\\
Last but not least, I owe a debt of gratitude to my parents and sisters, who provided me with the opportunity to study abroad, and endless love.

\cleardoublepage
\pagestyle{headings}
\chapter*{Abstract}
\addcontentsline{toc}{chapter}{Abstract}
\vspace{-1em}
Artificial intelligence and semantic technologies are evolving and have been applied in various research areas, including the education domain. Higher Education institutions (HEIs) strive to improve students' academic performance. Early intervention to at-risk students and a reasonable curriculum is vital for students' success. Prior research opted for deploying traditional machine learning models to predict students' performance. However, the existing research on applying deep learning models on prediction is very limited. In terms of curriculum semantic analysis, after conducting a comprehensive systematic review regarding the use of semantic technologies in the context of Computer Science curriculum, a major finding of the study is that technologies used to measure similarity have limitations in terms of accuracy and ambiguity in the representation of concepts, courses, etc. To fill these gaps, in this study, three implementations were developed, that is, to predict students’ performance using marks from the previous semester, to model a course representation in a semantic way and compute the similarity, and to identify the prerequisite between two similar courses. Regarding performance prediction, we used the combination of Genetic Algorithm and Long-Short Term Memory (LSTM) on a dataset from a Brazilian university containing 248730 records. As for similarity measurement between courses, we deployed Bidirectional Encoder Representation with Transformers (BERT) to encode the sentence in the course description from the Australian National University (ANU). We then used cosine similarity to obtain the distance between courses. With respect to prerequisite identification, TextRazor was applied to extract concepts from course descriptions, followed by employing SemRefD to measure the degree of prerequisite between two concepts. The outcomes of this study can be summarized as: (i) a breakthrough result improves Manrique’s work by 2.5\% in terms of accuracy in dropout prediction; (ii) uncover the similarity between courses based on course description; (iii) identify the prerequisite over three compulsory courses of School of Computing at ANU. In the future, these technologies could potentially be used to identify at-risk students, analyze the curriculum of university programs, aid student advisors, and create recommendation systems.
\\
\\
\textbf{Keywords: Dropout Prediction, Curriculum Semantic Analysis, Similarity Measurement, Prerequisite Identification, Genetic Algorithm, Long-Short Term Memory, Bidirectional Encoder Representation with Transformers, SemRefD}

\cleardoublepage
\microtypesetup{protrusion=false}  
\pagestyle{headings}
\markboth{Contents}{Contents}
\tableofcontents
\listoffigures
\listoftables
\microtypesetup{protrusion=true}

\mainmatter

\chapter{Introduction}
\label{cha:intro}


\section{Problem Statement and Motivations}
\label{sec:motivations}
Students' academic performance is not only important for themselves, but also important for higher education institutions (HEIs), who use it as a basis for measuring the success of their educational programs. A variety of metrics can be used to measure student performance. Dropout rates are notable among these metrics. In other words, the reduction of dropout rates could indicate the improvement of student's academic performance. However, according to the published figures, each year, one-fifth of first-year undergraduates drop out of their degrees across Australia \citep{shipley2019}. In Brazil, it is estimated that only 62.4\% of university enrolments succeed in obtaining an undergraduate degree \citep{Sales2016}. These are concerning statistics for a country's development and can affect students' lives financially, mentally, and professionally. As a consequence, HEIs and researchers have shown an increasing interest in predicting systems to identify students at risk of dropping out \citep{Ruben2019}. For example, Ruben et al. created multiple feature sets from student data to predict whether a student would drop out by using machine learning algorithms \citep{Ruben2019}; also, Mujica et al. pointed out that path analysis was useful for predicting student dropouts based on affective and cognitive variables \citep{Mujica2019}.

According to Vergel et al., the curriculum design may affect the students' performance and retention. A careful design is required to "understand the role of curriculum design in dropout rates and unmask how the political dynamics embedded in the curriculum influence students’ retention" \citep{Vergel2018}. Moreover, Drennan, Rohde, Johnson, and Kuennen claimed that students’ academic performance in a course is related to their performance in their prerequisite courses \citep{Drennan2002}, \citep{Johnson2006}. It can be seen that curriculum plays a crucial role in student performance as well as decreasing dropout rates.

The main motivation for this research is to improve students' academic performance based on the analysis of curriculum and student academic performance in different courses. Identifying prerequisites for a course is key to enhancing the student experience, improving curriculum design, and increasing graduation rates.

\section{Objectives}

The main objective of this study is to apply Artificial Intelligence and Semantic technologies to analyse and allow changes that may improve student academic performance. 

The goal of this research project is threefold:
\begin{itemize}
    \item Predict students’ performance in university courses using grades from previous semesters;
    \item Model a course semantic representation and calculate the similarity among courses; and, finally,
    \item Identify the sequence between two similar courses.
\end{itemize} 

To accomplish these goals, a systematic review was carried out to identify what technologies are being used in CS curriculum.

\section{Contributions}
\label{sec:contribution}
The main contributions of this thesis can be divided into:
\begin{itemize}
    \item An approach for student dropout prediction using Genetic Algorithm (GA) and Long Short-Term Memory (LSTM);
    \item A comprehensive systematic review to understand how Semantic Web and Natural Language Processing technologies have been applied to curriculum analysis and design in Computer Science; and, finally,
    \item An analysis of a Computer Science curriculum using the SemRefD distance proposed by Manrique et al. \citep{Ruben2019} and BERT proposed by Devlin et al. \citep{Jacob2018}.

\end{itemize}

\section{Thesis Outline}
\label{sec:outline}

Section 2 presents the background information on dropout prediction and text analysis. Section 3 covers the methodology used in this work including a systematic review and general techniques. Section 4 presents the experiments, results and discussion and, finally, section 5 concludes this thesis with future work directions.

\setcounter{tocdepth}{3}
\setcounter{secnumdepth}{3}

\chapter{Background and Related Work}
\label{cha:background}
The chapter is divided into two sections: background and related work. The following section briefly introduces the basic concepts unitized in this research. In related work, we conduct a systematic review regarding curriculum analysis by using Artificial Intelligence and Semantic Web technologies.
\section{Background}
Artificial Intelligence refers to the development of machines or computers that simulate or emulate the functions of the human brain. The functions of a computer differ according to the area of study. Prolog, for example, is a programming language that aims to understand human logic. It also applies mathematics in order to create systems that can discern relevant conclusions from a set of statements. Intelligent Agents is another example. Unlike traditional agents, Intelligent Agents are designed to take actions that are optimized to achieve a specific goal. Based on their perception of the environment and internal rules, intelligent agents make decisions. The study explores four growing fields: Machine Learning, Deep Learning, Natural Language Processing, and Semantic Web.

Machine Learning is a field of study that uses algorithms to detect patterns in data and to predict or make useful decisions. Different machine learning algorithms can be implemented in an innumerable number of scenarios. Machine learning algorithms are optimized for specific data sets; There are a variety of algorithms used to build models to suit different use cases. Using machine learning algorithms, we predict students' performance in university courses based on their performance in previous courses. 

Deep Learning is a neural network with multiple layers of perceptrons. The neural networks try to imitate human brain activities, albeit far from matching its capabilities, which enables it to study from high volume of data. Additional hidden layers can help to refine and optimize a neural network for accuracy, even when it has a single layer. Recurrent Neural Networks (RNNs) are one special type of neural networks that combines the information from previous time steps to generate updated outputs.

The Natural Language Processing (NLP) process is a psychological procedure that consists of analyzing successful people's strategies and applying them to reaching a personal goal. The process links cognitions, language, and patterns of action that are learned through experience to particular outcomes.

W3C (World Wide Web Consortium) standards enable the Semantic Web to be extended from the World Wide Web. W3C standardizes the process of making information on the World Wide Web machine-readable. In order to accomplish this goal, a series of communication standards were created that enable developers to describe concepts and entities, as well as their classification and relationships. The Resource Description Framework (RDF) and Web Ontology Language (OWL) have enabled developers to create systems that store and use complex knowledge data bases known as knowledge graphs. 

\section{Related Work}
This section discusses tools, frameworks, datasets, semantic technologies, and approaches used for curriculum design in Computer Science. Note that we present the methodology used to carry out a systematic review, but for brevity and adequacy we only present the most relevant works and sections. The complete systematic review will be submitted to a conference in Computer Science in Education as an outcome of this thesis.

\subsection{Methodology}
\label{sec:systematic_review}

\begin{figure}[h]
  \centering
  \includegraphics[width=0.5\textwidth]{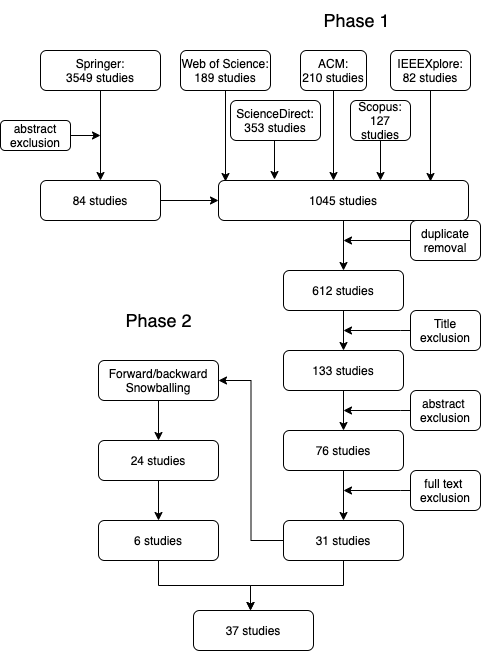}
  \caption{Paper Selection Process}
  \label{img1}
\end{figure}

This systematic review was conducted following the methodology defined by Kitchenham and Charters \citep{Kitchenham2007}. The method is composed of three main steps: planning, conducting and reporting.

The planning stage helps to identify existing research in the area of interest as well as build the research question. Our research question was created based on the PICOC method \citep{Kitchenham2007} and used to identify keywords and corresponding synonyms to build the search string to find relevant related works. The resulting search string is given below.

\emph{("computer science" OR "computer engineering" OR "informatics") AND ("curriculum" OR "course description" OR "learning outcomes" OR "curricula" OR “learning objects”) AND ("semantic" OR "ontology" OR "linked data" OR "linked open data")}

To include a paper in this systematic review, we defined the following inclusion criteria: (1) papers must have been written in English; (2) papers must be exclusively related to semantic technologies, computer science and curriculum; (3) papers must have 4 or more pages (i.e., full research papers); (4) papers must be accessible online; and, finally, (5) papers must be scientifically sound, present a clear methodology and conduct a proper evaluation for the proposed method, tool or model.

 

Figure \ref{img1} shows the paper selection process in detail. Initially, 4,510 papers were retrieved in total. We used the ACM Digital Library\footnote{\url{https://dl.acm.org/}}, IEEE Xplore Digital Library\footnote{\url{https://ieeexplore.ieee.org/}}, Springer\footnote{\url{https://www.springer.com/}}, Scopus\footnote{\url{https://www.scopus.com/}}, ScienceDirect\footnote{\url{https://www.sciencedirect.com/}} and Web of Science\footnote{\url{https://www.webofscience.com/}} as digital libraries. The Springer digital library returned a total of 3,549 studies. The large number of papers returned by the Springer search mechanism led us to develop a simple crawling tool to help us in the process of including/rejecting papers in our systematic review. All the information returned by Springer was collected and stored in a relational database. After that, we were able to correctly query the Springer database and select the relevant papers for this systematic review. 

We also applied the forward and backward snowballing method to this systematic review to identify relevant papers that were not retrieved by our query string. The backward snowballing method was used to collect relevant papers from the references of the final list of papers in Phase 1, whereas the forward snowballing method was used to collect the papers that cited these papers \citep{Wohlin2014}. Google Scholar\footnote{\url{https://scholar.google.com/}} was used in the forward snowballing method. In total, 37 studies were identified as relevant; most of the studies were published in the last few years, which shows the increasing relevancy of this topic.

In the following sections we present the most relevant works that inspired the proposed approaches reported in this thesis.







\subsection{Tools/Frameworks}


Protégé\footnote{\url{https://protege.stanford.edu/}}\citep{Imran2016} is a popular open-source editor and framework for ontology construction, which several researchers have adopted to design curricula in Computer Science \citep{Tang2013}, \citep{Wang2019}, \citep{Nuntawong2017}, \citep{Karunananda2012}, \citep{Hedayati2016}, \citep{Saquicela2018SimilarityDA}, \citep{Maffei2016}, and \citep{Vaquero2009}. Despite its wide adoption, Protégé still presents limitations in terms of the manipulation of ontological knowledge \citep{Tang2013}. As an attempt to overcome this shortcoming, Asoke et al. \citep{Karunananda2012} developed a curriculum design plug-in called OntoCD. OntoCD allows curriculum designers to customise curricula by loading a skeleton curriculum and a benchmark domain ontology. The evaluation of OntoCD is performed by developing a Computer Science degree program using benchmark domain ontologies developed in accordance with the guidelines provided by IEEE and ACM. Moreover, Adelina and Jason \citep{Tang2013} used Protégé to develop SUCO (Sunway University Computing Ontology), an ontology-specific Application Programming Interface (API) for curricula management system. They claim that, in response to the shortcoming with using the Protégé platform, SUCO shows a higher level of ability to manipulate and extract knowledge and will function effectively if the ontology is processed as an eXtensible Markup Language  (XML) format document.

Other specific ontology-based tools in curriculum management have also been developed. CDIO \citep{Liang2012} is an example of such tool. CDIO was created to automatically adapt a given curriculum according to teaching objectives and teaching content based on the designed constructed ontology. A similar approach was used by Maffei et al. /citep{Maffei2016} to model the semantics behind functional alignments in order to design, synthesize, and evaluate functional alignment activities. In Mandić's study, the author presented a software platform\footnote{\url{http://www.pef.uns.ac.rs/InformaticsTeacherEducationCurriculum}} for comparing chosen Curriculum for information technology teachers \citep{Mandic2018}. In Hedayati's work, the authors used the curriculum Capability Maturity Model (CMM), which is a taxonomical model used for describing the organization’s level of capability in the domain of software engineering \citep{Paulk1993}, An ontology-driven model for analyzing the development process of the vocational ICT curriculum in the context of the culturally sensitive curriculum in Afghanistan is used as a reference model \citep{Hedayati2016}.

\subsection{Datasets}
There are several datasets used in curriculum design studies in Computer Science. The open-access CS2013\footnote{\url{https://cs2013.org/}} dataset is the result of the joint development of a computing curriculum sponsored by the ACM and IEEE Computer Society \citep{Piedra2018}. The CS2013 dataset has been used in several studies \citep{Piedra2018}, \citep{Aeiad2016}, \citep{Nuntawong2016}, \citep{Nuntawong2017}, \citep{Karunananda2012}, \citep{Hedayati2016}, and \citep{Fiallos2018} to develop ontologies or as a benchmark curriculum in similarity comparison between computer science curricula.


Similar to CS2013, The Thailand Qualification Framework for Higher Education (TQF: HEd) was developed by the Office of the Thailand Higher Education Commission to be used by all higher education institutions (HEIs) in Thailand as a framework to enhance the quality of course curricula, including the Computer Science curriculum. TQF: HEd was used for the guidelines in terms of ontology development in the following studies \citep{Nuntawong2017}, \citep{Nuntawong2016}, \citep{Hao2008}, and \citep{Nuntawong2015}.

Other studies use self-created datasets (e.g., \citep{Wang2019}, \citep{Maffei2016}, \citep{Hedayati2016},  and \citep{Fiallos2018}). Specifically, in Wang's work, the Ontology System for the Computer Course Architecture (OSCCA) was proposed based on a dataset created using course catalogs from top universities in China as well as network education websites \citep{Wang2019}. In Maffei's study, the authors experimented and evaluate the proposal based on the Engineering program at KTH Royal Institute of Technology in Stockholm, Sweden \citep{Maffei2016}. In Gubervic et al.'s work, the dataset used in comparing courses comes from Faculty of Electrical Engineering and Computing compared to all universities from United States of America \citep{Guberovic2018}. In Fiallos's study, not only did the author 
adopt CS2013 for domain ontologies modeling, but also the core courses from Escuela Superior Politécnica del Litoral (ESPOL\footnote{\url{https://www.espol.edu.ec/}}) Computational Sciences were collected for semantic similarity comparison \citep{Fiallos2018}.



\subsection{Languages, Classes and Vocabulary}
RDF is used as the design standard for data interchange in the following studies \citep{Piedra2018}, \citep{Nuntawong2017}, and \citep{Saquicela2018SimilarityDA}. In particular, Saquicela et al. \citep{Saquicela2018SimilarityDA} generated curriculum data in the RDF format, creating and storing data in a repository when the ontological model has been defined and created.

OWL is an extension of RDF that adds additional vocabulary and semantics to the classic framework \citep{McGuinness2004}. OWL is used in many studies \citep{Piedra2018}, \citep{Adrian2020}, \citep{Mandic2018}, \citep{Wang2019}, \citep{Maffei2016}, and \citep{Vaquero2009} for representing and sharing knowledge on the Web\footnote{\url{https://www.w3.org/OWL/}}. Apart from OWL, only two studies used XML\footnote{\url{https://www.w3.org/standards/xml/}} due to implementation requirements of research (\citep{Tang2013} and \citep{Hao2008}).

Body of Knowledge (BoK), a subclass of OWL, is a complete set of concepts, terms and activities, which can represent the accepted ontology for a professional domain \citep{Piedra2018}. BoK has become a common development method in many studies \citep{Piedra2018}, \citep{Nuntawong2017}, \citep{Nuntawong2016}, \citep{Hao2008}, \citep{Karunananda2012}, \citep{Tapia2018}, \citep{Chung2014}, and \citep{Nuntawong2015}. In Piedra's study \citep{Piedra2018}, the BoK defined in CS2013 ontology was viewed as a to-be-covered description of the content and a curriculum to implement this information. Similarly, Numtawong et al.\citep{Nuntawong2017} applied the BoK, which is based on the ontology of TQF: HEd, to conduct the ontology mapping.

The Library of Congress Subject Headings\footnote{\url{https://www.loc.gov/aba/cataloging/subject/}} (LCSH) is a managed vocabulary Upheld by the Library of Congress. LCSH terminology is a BoK, which contains more than 240,000 topical subject headings. The equivalence, hierarchical, and associative types of relationships between headings can be offered. In Adrian's study, the authors create an ontology based on LCSH and the Faceted Application of Subject Terminology\footnote{\url{https://www.oclc.org/en/fast.html}} (a completely enumerative faceted subject terminology schema originated from LCSH), to assess the consistency of an academic curriculum and apply it to an Information Science curriculum \citep{Adrian2020}.

Knowledge Area (KA), also a subclass of OWL, is an area of specialization such as Operating Systems and Algorithm. The relationship between BoK and KA is built in various ways in the studies. For example, in Piedra's study \citep{Piedra2018}, each BoK class contains a set of KAs. In contrast, Numtawong et al. \citep{Nuntawong2017} considered KA as the superclass of BoK. KA classification was proposed in Orellana et al.'s study \citep{Orellana2018}. In that paper, the curricula are classified in the KA defined by UNESCO\footnote{\url{https://whc.unesco.org/en/}} which defines 9 main areas and 24 subareas of knowledge. To do this, they convert the curricula to the vector space and then process with using traditional supervised approaches such as the support vector machines and k-nearest neighbors \citep{Orellana2018}. By classifying the KA, the similarity measurement can be applied more easily.

\subsubsection{Curriculum Design Analysis and Approaches}
One development approach found is the extraction and interrelationship analysis. NLTK\footnote{\url{https://www.nltk.org/}} is deployed to segment raw data into terms, using various algorithms to extract and analyse the interrelationship of items, then construct an ontology by using a certain framework such as Protégé as in \citep{Piedra2018}, \citep{Wang2019}, and \citep{Tapia2018}.

Text Mining, also known as text analysis methods, have been used to find the interesting patterns in HEIs' curriculum \citep{Orellana2018}.
With using Text Mining approaches, keywords can be extracted from both found documents and course materials for the further comparison and analysis \citep{Kawintiranon2016}. In the next section, the application of Text Mining approaches in Computer Science curriculum will be elaborated.



In the context of curriculum similarity measurement, Gomaa and Fahmy \citep{H.Gomaa2013} define string-based, corpus-based, knowledge-based and hybrid similarity measures. String-based measures the distance between string (words) that can be compared by characters or terms. Corpus-based approach measures the semantic meaning of terms and phrases which are provided in the corpus. Knowledge-based uses synsets-formed word networks such as WordNet to compare cognitive meaning between each other. String-based similarity was conducted by Corpus-based was done in \citep{Orellana2018}, \citep{Pawar2018}, \citep{Aeiad2016} and \citep{Fiallos2018}. Knowledge-based similarity approach is proposed in \citep{Nuntawong2015}.

String-based similarity between terms was measured in many studies \citep{Orellana2018}, \citep{Pawar2018}, \citep{Adrian2020}, \citep{Seidel2020}, \citep{Wang2019}, and \citep{Saquicela2018SimilarityDA}. Orellana et al. \citep{Orellana2018} used cosine similarity between terms to acquire the level of similarity between two course descriptions. Adrian and Wang used the same approach to measure the similarity \citep{Adrian2020}, \citep{Wang2019}. 

Pawar and Mago utilized the Bloom's taxonomy to measure the similarity of sentence pairs in the Learning outcomes \citep{Pawar2018}. In another paper, Saquicela et al. used the K-means, an unsupervised clustering algorithm to calculate the similarity among courses content\citep{Saquicela2018SimilarityDA}.

Corpus-Based similarity measurement was proposed in these studies \citep{Orellana2018}, \citep{Pawar2018}, \citep{Adrian2020}, and \citep{Fiallos2018}. In Orellana et al.'s study, the topics are extracted and processed by LSA. Through the process, the terms and documents are located within their context to get the most relevant documents (Wikipedia articles and curricula) by adding the similarity threshold.

Knowledge-based (Semantic) similarity measurement was proposed in \citep{Nuntawong2016} and \citep{Pawar2018}. In Nuntawong's paper \citep{Nuntawong2017}, the authors designed curriculum ontology and defined the ontology mapping rules by semantic relationships that can be established between curricula. After completing the steps above, an ontological system was built by converting input curriculum data to ontology. Retrieve BoK in KA from TQF: HEd and course descriptions which were compared to WordNet. In the end, calculate the semantic similarity values with using extended Wu \& P's algorithm \citep{Wu1994}. To calculate the semantic similarity between words, Pawar and Mago used Synsets from WordNet. The method simulates supervised learning through the use of corpora. Additionally, they used the NLTK-implemented max similarity algorithm to determine the sense of the words \citep{Pawar2018}.

One important component in curricula is the Learning Outcomes (LOs). It defines what students are expected to learn by taking the course. There are six layers in a hierarchical structure in Bloom's Taxonomy (Remembering, Understanding, Applying, Analysing, Evaluating, and Creating) \citep{Lasley2013}. Semantic technologies are used along with the Bloom Taxonomy to calculate the similarity of learning outcomes between courses. Pawar and Mago \citep{Pawar2018} propose a semantic similarity metric using WordNet\footnote{\url{https://wordnet.princeton.edu/}} to generate a score to compare LOs based on the Bloom taxonomy. Similarly, Mandi\'c \citep{Mandic2018} proposed the taxonomic structural similarity between curricula by applying the revised Bloom's taxonomy which has the adjustment in cognitive levels of learning.


This section partially presented some of the works found in the literature that used semantic technologies to help university stakeholders design curricula for Computer Science. The following chapters present the approaches and analysis carried out using previous works as inspiration.

\chapter{Approaches}
\label{cha:methodology}
\setcounter{tocdepth}{3}
\setcounter{secnumdepth}{3}
This chapter contains two sections. The first section introduces the dropout prediction task and a series of techniques to predict students' performance. The second part presents semantic techniques that can be used to analyze the relationship between Computer Science courses.

\section{Dropout Prediction}
\label{sec:dropout_prediction}
This section presents an approach to predicting dropout. We illustrate the entire dropout prediction workflow followed by the description of the dataset used and the corresponding pre-processing measurements. We then present an SVM-based genetic algorithm (GA) for feature selection and our LSTM approach for dropout prediction.

\subsection{Procedure Overview}
\label{subsec:procedure_overview}
\begin{figure}
  \centering
  \includegraphics[width=0.9\textwidth]{{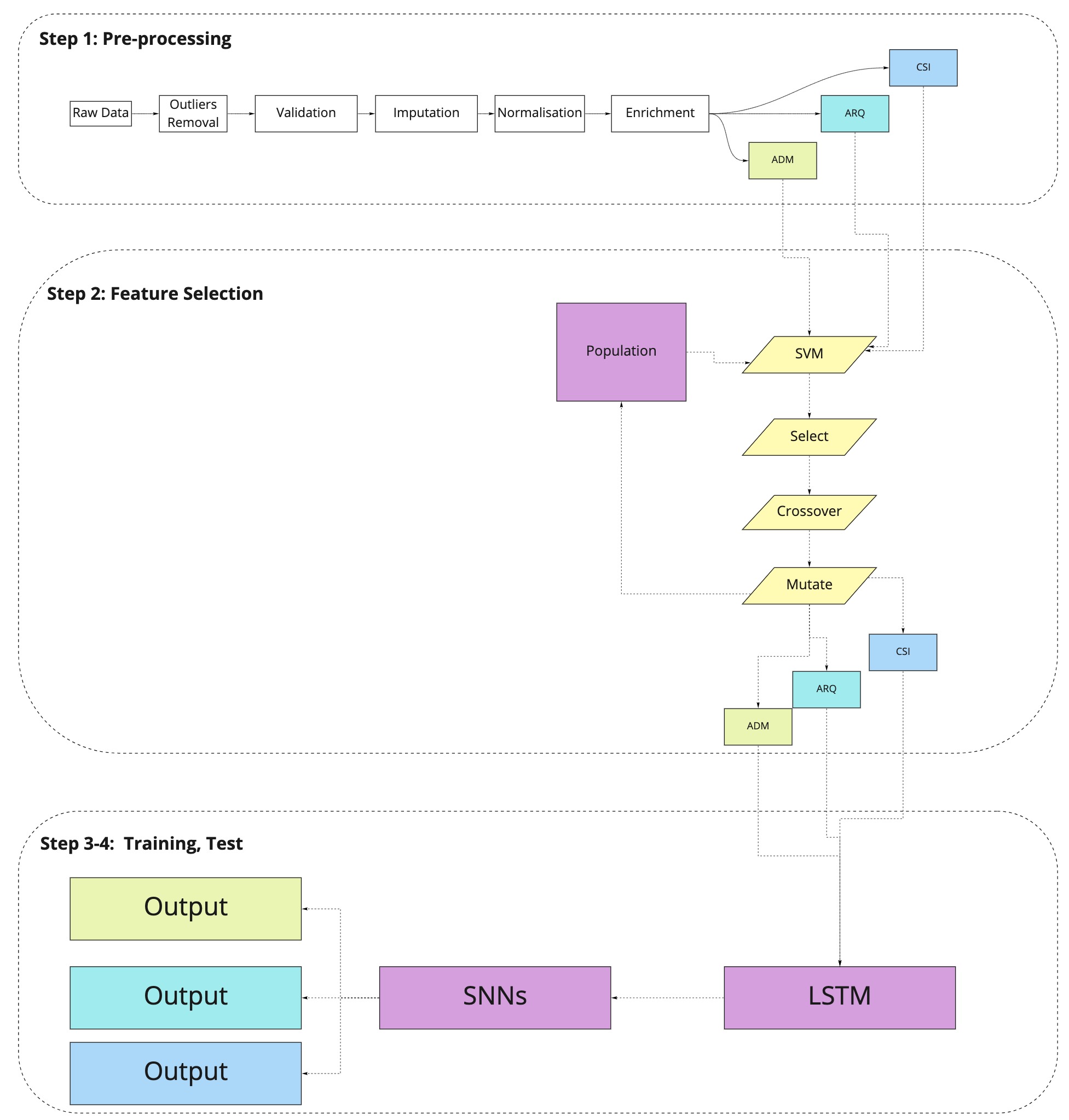}}
  \caption{The workflow of dropout prediction}
  \label{fig:entire_dp_workflow}
\end{figure}

As illustrated in Figure~\ref{fig:entire_dp_workflow}, the entire dropout prediction workflow is composed of four steps. Details of each step will be explained in the following sections.



Briefly, the first step of dropout prediction is responsible for data pre-prepocessing including obtaining datasets. This step uses data wrangling and machine learning (ML) techniques. The second step is responsible for the feature selection. It uses the pre-processed data outputted in the first step. Steps 3 and 4 are merged together and start with training and testing the Long Short-term Memory (LSTM) and Fully Connected (FC) neural network.



\subsection{Data Pre-processing}
\label{subsec:data_and_pre-processing}
\begin{figure}
  \centering
  \includegraphics[width=0.9\textwidth]{{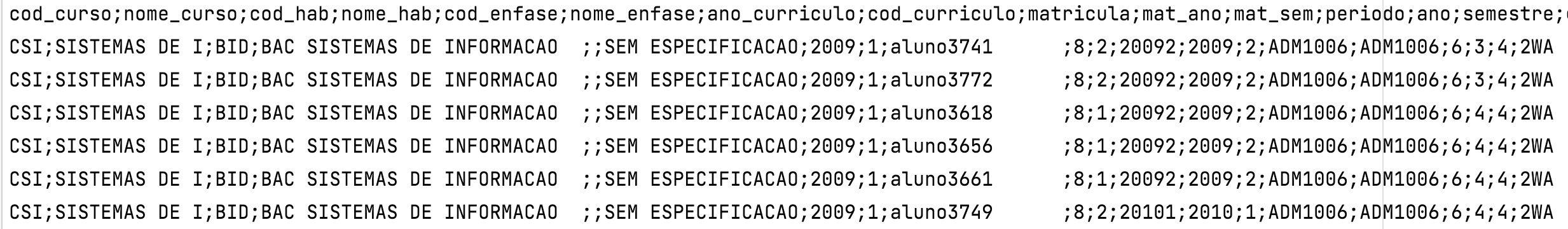}}
  \caption{Dataset snippet}
  \label{fig:dataset}
\end{figure}

Before introducing the data pre-processing, we introduce the data set used for this experiment. Figure \ref{fig:dataset} presents a few instances of the dataset used. 



\begin{figure}[H]
\centering
\begin{subfigure}{.5\textwidth}
  \centering
  \includegraphics[width=\linewidth]{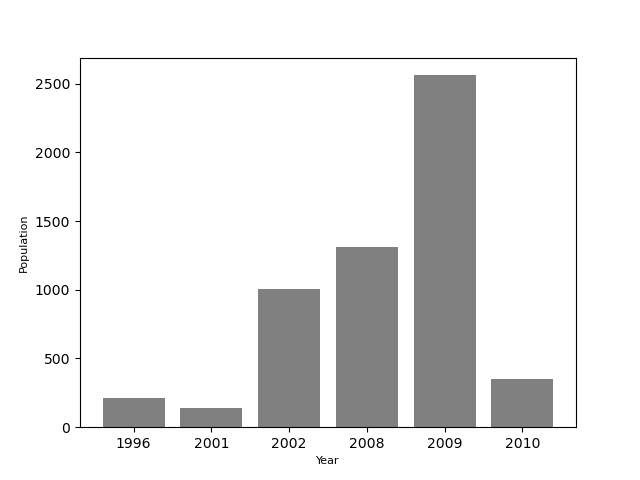}
  \caption{Student distribution by year}
  \label{fig:year_distribution}
\end{subfigure}%
\begin{subfigure}{.5\textwidth}
  \centering
  \includegraphics[width=\linewidth]{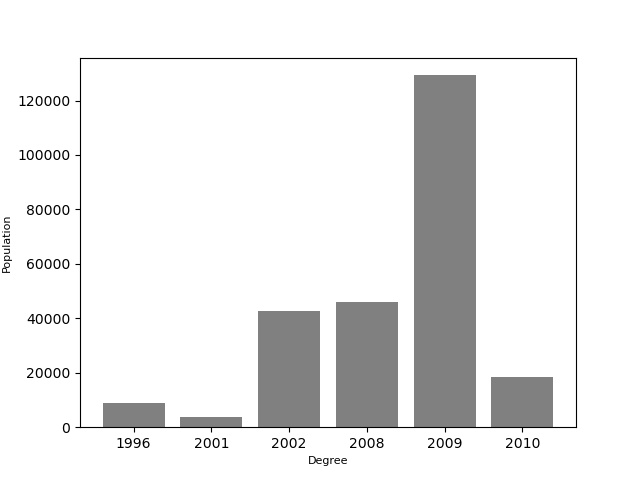}
  \caption{Record distribution by year}
  \label{fig:record_distribution}
\end{subfigure}
\caption{Distribution by year}
\label{fig:year_distribution}
\end{figure}

\begin{figure}[H]
\centering
\begin{minipage}{.5\textwidth}
  \centering
  \includegraphics[width=\linewidth]{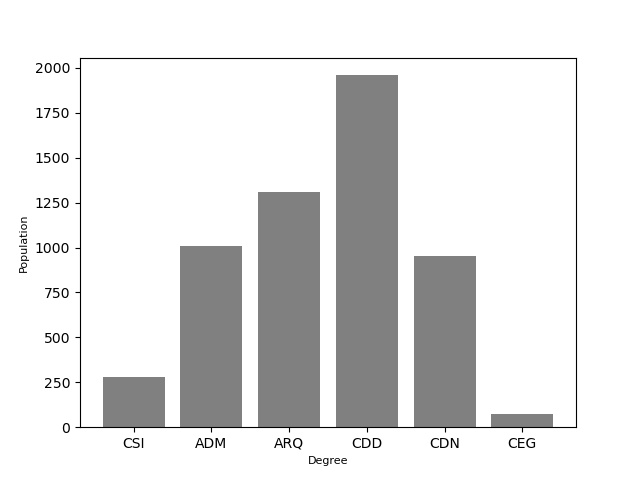}
  \captionof{figure}{Student distribution by degree}
  \label{fig:degree_distribution}
\end{minipage}%
\begin{minipage}{.5\textwidth}
  \centering
  \includegraphics[width=\linewidth]{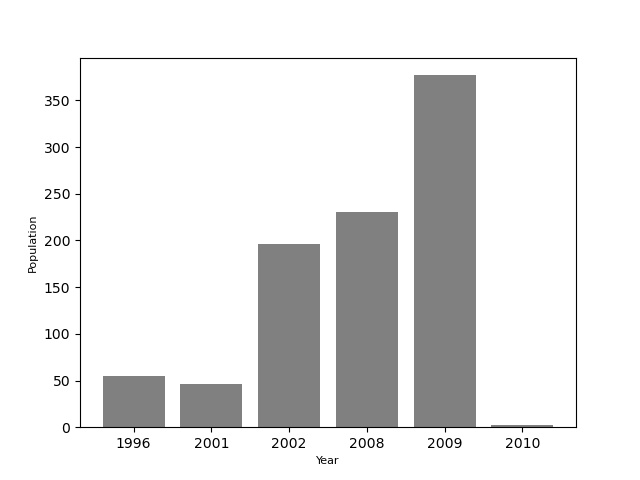}
  \captionof{figure}{Dropout distribution by year}
  \label{fig:dropout_distribution}
\end{minipage}
\end{figure}


The dataset used in this thesis has been used in previous research \citet{Manrique2019} and is provided by a Brazilian university. It contains 248,730 academic records of 5,582 students enrolled in six distinct degrees from 2001 to 2009. The dataset is in Portuguese and we will translate the main terms for better understanding. 

The dataset contains 32 attributes in total: cod\textunderscore curso (course code); nome\textunderscore curso (course name); cod\textunderscore hab (degree code); nome\textunderscore hab (degree name); cod\textunderscore enfase (emphasis code); nome\textunderscore enfase (emphasis name); ano\textunderscore curriculo (curriculum year); cod\textunderscore curriculo (curriculum code); matricula (student identifier); mat\textunderscore ano (student enrolment year); mat\textunderscore sem ((semester of the enrolment); periodo (term); ano (year); semestre (semester); grupos (group); disciplina (discipline/course); semestre\textunderscore recomendado (recommended semester); semestre\textunderscore do\textunderscore aluno (student semester); no\textunderscore creditos (number of credits); turma (class); grau (grades); sit\textunderscore final (final status (pass/fail)); sit\textunderscore vinculo\textunderscore atual (current status); nome\textunderscore professor (professor name); cep (zip code); pontos\textunderscore enem (national exam marks); diff (difference between semesters and students performance); tentativas (number of attempts in a course); cant (previous course); count (count); identificador (identifier); nome\textunderscore disciplina (course name). 

Most of the data is related to the status of a student in a give course and degree. The "semestre" attribute is related to the semester a student took a course and has been previously used to create a time series of students \citet{Ruben2019}.
The attribute "sit\textunderscore vinculo\textunderscore atual" indicates that there are 12 status of enrolment where three of them ("DESLIGADO","MATRICULA EM ABANDONO","JUBILADO") represent the dropout status. Grades is scaled from 0 to 10 inclusive where 10 is the maximum grade. The dataset is anonymous and the identifiers do not allow re-identification. The students' ids are encoded with dummy alphanumeric code such as "aluno1010" ("student1010"). The dataset was profiled by year and degree and is presented in Figure \ref{fig:year_distribution} - \ref{fig:dropout_distribution}.

\begin{table*}
  \centering
  \caption{Data wrangling techniques used in pre-processing.}
  \label{tab:preprocess}
  \resizebox{0.7\textwidth}{!}{
\begin{tabular}{BCc^c}
\toprule
\rowstyle{\bfseries}
Technique     & Operation         \\
\midrule 
Data Cleaning       & outliers/duplicates correction            \\
Data Validation     & inappropriate data identification                 \\
Data Enrichment     & Data enhancement         \\
Data Imputation     & missing values filling        \\
Data Normalisation  & data re-scale    \\ 
\bottomrule
\end{tabular}
}
\end{table*}


To pre-process the data, we first split and reorganised the data set. We split the dataset into three parts: CSI ("Information Systems"), ADM ("Management"), and ARQ ("Architecture") datasets. After that we removed duplicate data, resolved data inconsistencies and removed outliers. For example, in the attribute "grades", a few marks were greater than 10, an invalid instance given the university marking scheme, and therefore removed. Attributes were also removed such as "grupos" as it was irrelevant and basically a copy of another attribute ("disciplinas"). Table \ref{tab:preprocess} presents all information of the data-preprocessing step.

Another technique used was data validation where inappropriate data is identified and removed. As part of the data pre-processing step, we also converted categorical to numerical data. For this step we used a tool named LabelEncoder which is a extensive library of Sklearn\footnote{\url{https://scikit-learn.org/}}.

For the dropout status attribute ("sit\textunderscore vinculo\textunderscore atual"), the instances were replaced by 0 (dropout) and 1 (enrolled). With respect to data imputation, the Random Forest (RF), which is a popular Machine Learning algorithm for constructing multiple decision trees in training, was employed to implement. As for normality or non-normality, linearity or non-linearity type of data, RF has a better performance than other algorithms in terms of handling imputation\citep{Pantanowitz2009}, \citep{Shah2014}, and \citep{Hong2020}. To be specific, the sequence of imputation is from the least column to the ones with more missing data, filling the missing values with 0 in other columns, and then running the RF and performing iterations until each missing value will be handled. The procedure of imputation is shown in Algorithm \ref{alg:imputation}. Next comes to group the data by "course" and "semester" attributes.




The final step normalizes all the input data except the "sit\textunderscore vinculo\textunderscore atual" and "semestre" with the \emph{z}-score method as \emph{z}-score can improve the model performance than other techniques \citep{Imron2020}, and \citep{Chris2003}. The \emph{z}-score method is computed based on a set of values \begin{math}\mathit{S}=\left \{ x_{1},x_{2},...,x_{n} \right \}\end{math}, the mean of the set (\begin{math}\mu \end{math}), and the standard deviation (\begin{math}\sigma\end{math}). The formula is presented in what follows.

\begin{equation}
    \emph{z}=\frac{\mathit{x}-\mu}{\sigma }
    \label{for: z-zeros}
\end{equation}

The last pre-processing step applied to the dataset it the data enrichment step. As Table \ref{tab:preprocess} shows, the number of instances in the CSI dataset is very small and not sufficient for training/testing purposes. For this, we used the Synthetic Minority Oversampling Technique (SMOTE) to increase the number of instances of the CSI dataset. Briefly, SMOTE generates new instances based on real ones. SMOTE considers the minorities to generate and balance the dataset with new instances.

\algdef{SE}[DOWHILE]{Do}{doWhile}{\algorithmicdo}[1]{\algorithmicwhile\ #1}%
\begin{algorithm}
\caption{Procedure of Imputation}
\hspace*{\algorithmicindent} \textbf{Input}: Dataset with missing values \\
\hspace*{\algorithmicindent} \textbf{Output}: Complete dataset with predicted values
\begin{algorithmic}[1]
\Procedure{}{}
    \State Get the order of imputation column by the number of missing values
    \Do
        \State Fill NULL with 0 in all NULL except the column which has least number of missing values\
        \State Train RF to predict the missing values\
    \doWhile{Number of missing values > 0} \Comment{Perform Iteration again}
\EndProcedure
\end{algorithmic}
\label{alg:imputation}
\end{algorithm}

\subsection{Feature Selection}
\label{subsec:feature_selection}

After the data is pre-processed, to reduce the time consumption and increase the computational efficiency in training, it is necessary to apply feature selection to remove irrelevant features. Consequently, Support Vector Machine(SVM)-based Genetic Algorithm (GA) will be introduced in this section.

\begin{figure}
  \centering
  \includegraphics[width=0.6\textwidth]{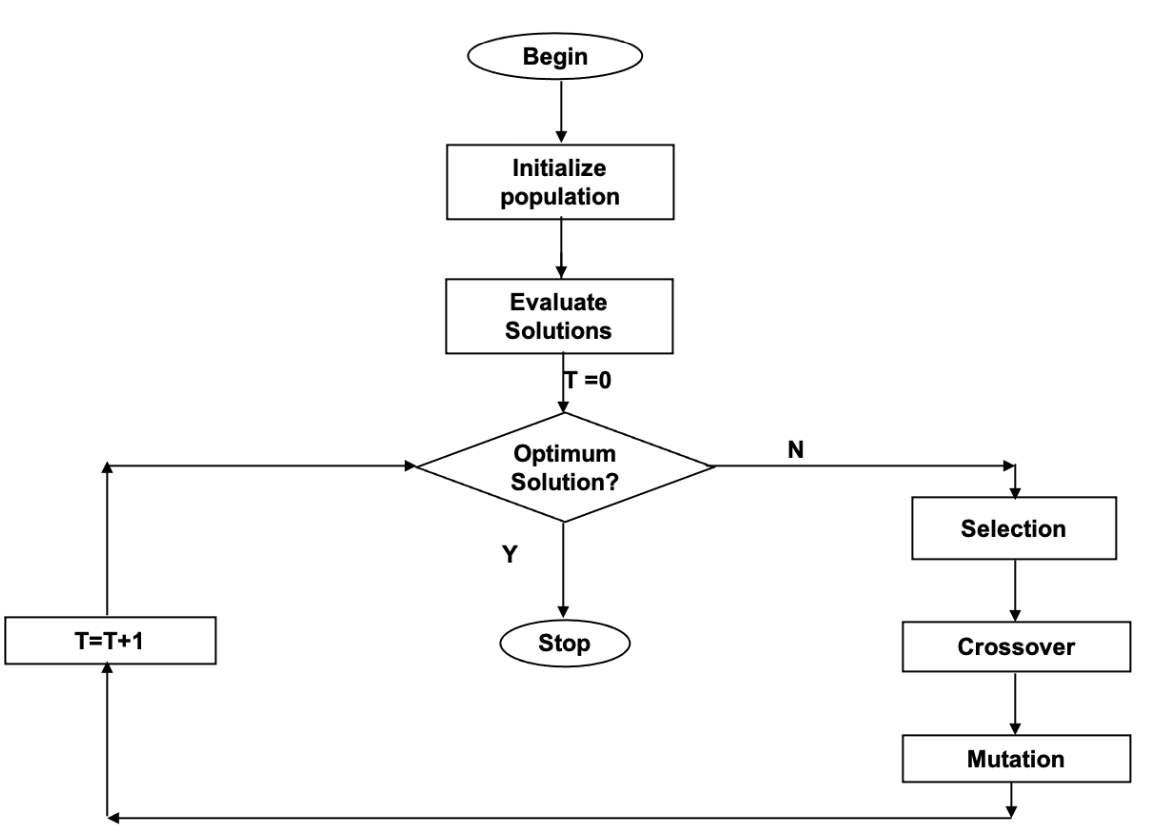}
  \caption{Procedure of GA}
  \label{fig:ga}
\end{figure}

In GA, optimization strategies are implemented based on the simulation of evolution by natural selection for a species. Table /ref{tab:ga} lists popular GA operators for creating optimal solutions and solving search problems based on biologically inspired principles such as mutation, crossover, and selection. Moreover, as its nature mentioned previously, the application of GA for feature selection is popular and was proven to improve the model performance in \citep{Babatunde2014}, \citep{Huang2007}, and \citep{Leardi2000}. The whole procedure is displayed in Figure \ref{fig:ga}. In this thesis, to perform a GA-based feature selection, a set of techniques was deployed. First, a population was randomly generated as the initial solution whose size is 1,000, followed by binary encoding which determines the chosen feature as 1, and 0 represents the column that will not be chosen. The next step is to calculate the fitness of each individual in the population by performing SVM in the three datasets (ADM, CSI, and ARQ). SVM is a supervised linear machine learning technique that is most commonly employed for classification purposes, and it has good performance in the fitness calculation in feature selection \citep{TAO2019323}. It is defined as a linear classifier with a maximum interval on the feature space (when a linear kernel is used), which is basically an interval maximization strategy that results in a convex quadratic programming problem. Given a training dataset, \begin{math}\mathit{D}=\left \{ ([x_{1_{1}},x_{1_{2}},...x_{1_{n}}], y_{1}),([x_{2_{1}},x_{2_{2}},...x_{2_{n}}], y_{2}),...,([x_{m_{1}},x_{m_{2}},...x_{m_{n}}], y_{n}) \right \},y_{n}\in \{{-1,+1}\}\end{math}, where $\mathit{m}$ is the number of samples in the dataset, and $\mathit{n}$ is the number of features. In this experiment $n=27$. The target is to get a decision boundary so as to separate the samples into different areas. The separator in two-dimensional space is a straight line and can be written as formula \ref{for: 2-d}:

\begin{equation}
    \textit{m}\textit{x}+\textit{c}=0
    \label{for: 2-d}
\end{equation}

Once mapping the separator to the $n$-dimensional space, it will become the hyperplane separator $\mathit{H}$. It can be written as formula \ref{for: h}:

\begin{equation}
    \mathit{H}: \mathit{w}^{T}\textit{x} + b = 0
    \label{for: h}
\end{equation}

where $\mathit{w}= (\mathit{w_1},\mathit{w_2},...\mathit{w_{n}})$ where w is the vector determining the hyperplane direction, The hyperplane is located at the origin after a displacement term $\mathit{b}$ is applied. By determining the vector $\mathit{w}$ and the bias $\mathit{b}$, the division hyperplane is denoted as ($\mathit{w}$, $\mathit{b}$). In the sample space, a given point vector $\varnothing (\mathit{x_o})$ is the distance from the hyperplane. \ref{for: d}:

\begin{equation}
    \mathit{d}_{H}(\varnothing (\mathit{x_o})))=\frac{\left | \mathit{w}^{T}\varnothing (\mathit{x_o})+\mathit{b} \right |}{\left \| \mathit{w} \right \|_{2}}
    \label{for: d}
\end{equation}

where $\left \| \mathit{w} \right \|_{2}$ is the 2-norm which can be defined by $\mathit{w}$ as in formula \ref{for: 2-norm}:

\begin{equation}
    \left \| \mathit{w} \right \|_{2}=\sqrt{\mathit{w}_{1}^{2}+\mathit{w}_{2}^{2}+...+\mathit{w}_{n}^{2}}
    \label{for: 2-norm}
\end{equation}

Suppose hyperplane ($\mathit{w}$, $\mathit{b}$) can classify the training dataset, so as to $(x_{o},y_{o})\in \mathit{D}$, the following corollary can be obtained as below:
\begin{equation}
\left\{\begin{matrix}
 & \mathit{w}^{T}\varnothing (\mathit{x_o})+\mathit{b}\geq 1, \quad \mathit{y_m}=+1   \qquad \mathit{if \; correct}\\
 & \mathit{w}^{T}\varnothing (\mathit{x_o})+\mathit{b}\leq 1, \quad \mathit{y_m}=-1   \qquad \mathit{if \; incorrect}
\end{matrix}\right.
\label{for: a}
\end{equation}

The vectors close to the hyperplane make the formula \ref{for: a} hold are named "support vectors". The sum of the distance of 2 vectors which are from different areas to the hyperplane is called "margin", which is defined in the formula \ref{for: distance}:
\begin{equation}
\gamma =\frac{2}{\left \| \textit{w} \right \|_{2}}
\label{for: distance}
\end{equation}

To obtain a partitioned hyperplane with maximum margin, that is, to find parameters $\mathit{w}$ and $\mathit{b}$ that comply with the constraint in equation \ref{for: a} such that the optimal hyperplane classifies all the points in $\mathit{D}$ correctly, namely:

\begin{equation}
\mathit{w}^{*}=\mathit{arg}_{\mathit{w}}\mathit{max}(\gamma),\ \mathit{s.t.}min_{n}\mathit{y^{n}}\begin{bmatrix}
\mathit{w}^{T}\varnothing (\mathit{x_o})+\mathit{b} =1
\end{bmatrix}
\label{for : final}
\end{equation}

Followed by the fitness computation with SVM in 5-fold cross-validation, which is a technique to prevent overfitting, the proportional selection was conducted to get the individuals, the process is similar to the roulette wheel, that is, individuals with higher fitness ratings will have the greater chance to be selected, the formula of this process is shown in equation \ref{for: selection}:

\begin{equation}
\varphi _{s}(\mathit{x_{i}}(\mathit{t}))=\frac{\textit{f}_{\gamma }(\mathit{x_{i}}(\mathit{t}))}{ \sum_{\textit{I}=1}^{\mathit{n}}\textit{f}_{\gamma }(\mathit{x_{I}}(\mathit{t}))}
\label{for: selection}
\end{equation}

where $\mathit{n}_{s}$ is a population's total number of chromosomes whose initial number is 1,000, $\varphi _{s}(\mathit{x_{i}})$ is the possibility of $\mathit{x_{i}}$ being selected , and $\textit{f}_{\gamma }(\mathit{x_{i}})$ is the fitness of $\mathit{x_{i}}$ to yield float type of value greater than 0.

\begin{figure}
  \centering
  \includegraphics[width=0.6\textwidth]{{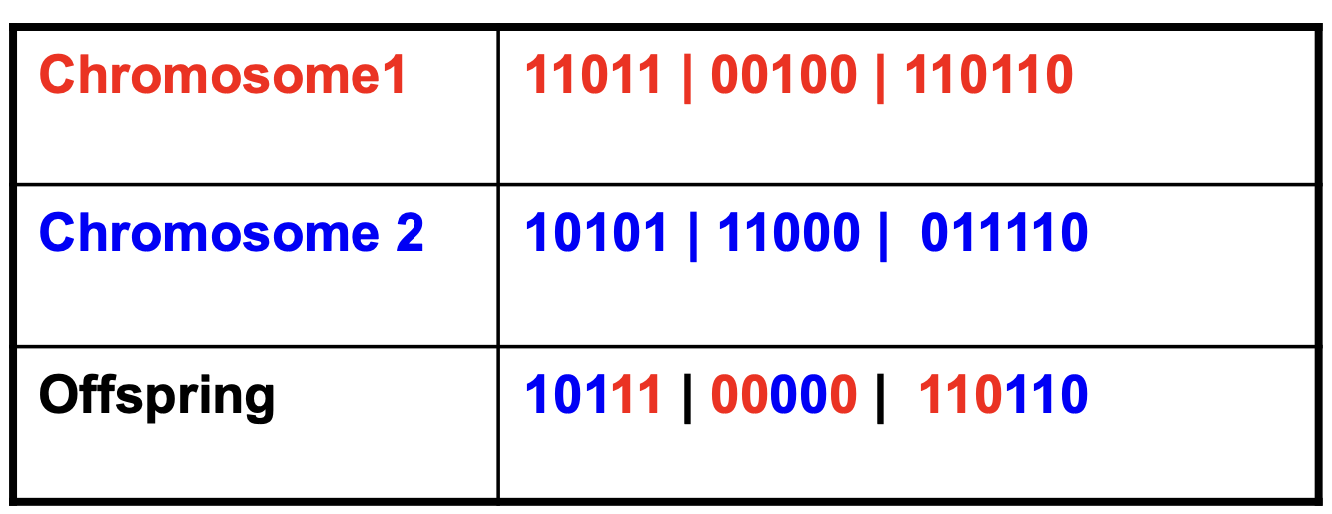}}
  \caption{Uniform Crossover}
  \label{fig:crossover}
\end{figure}

\begin{figure}
  \centering
  \includegraphics[width=0.6\textwidth]{{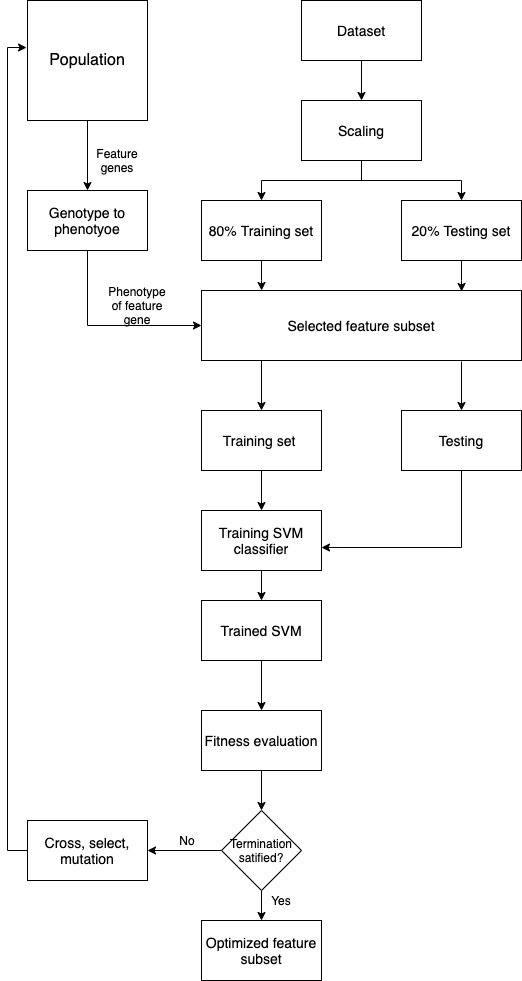}}
  \caption{The workflow of GA+SVM feature selection}
  \label{fig:GA}
\end{figure}

After completing the selection, the uniform crossover was performed to get the offspring between a pair of selected parents. Specifically, each gene is selected at random from one of the corresponding genes on every parent chromosome as shown in Figure \ref{fig:crossover}. Note that this method will only yield one offspring and the cross rate is 0.75. Reproduction is followed by mutation which certain gene(s) will be changed randomly by the setting rate, in this thesis, is 0.002. The final step is the terminal criterion, when the maximum generation is reached, the population has evolved 100 generations, and an optimal subset will be generated as the output. The entire workflow is illustrated in Figure \ref{fig:GA}.

\begin{table*}
  \centering

  \caption{Genetic Algorithm techniques, descriptions, goals, and specific method used in this study}
  
  \label{tab:ga}
  \resizebox{\textwidth}{!}{
\begin{tabular}{BCc^c^c^c}
\toprule
\rowstyle{\bfseries}
Operator            & Description & Goal    & Method deployed \\
\midrule 
Initialization      & initialization of population acquisition  & to generate a set of solutions                        & Randomly generated initialization\\
Encoding            &  representation of the individuals        & to convey the necessary information                   & Binary Encoding \\
Fitness             & the degree of health of individuals       & to evaluate the optimality of a solution               & Support Vector Machine (SVM)\\
Crossover           & parents are chosen for production         & to determine which solutions are to be preserved     & Uniform Crossover \\
Selection           & solutions are selected for production     & to select the individuals with better adaptability  & Proportional Selection  \\ 
Mutation            & a gene is deliberately changed            & to maintain diversity in the population set           & mutation rate setting-up\\
Termination         & a process stops the evolution             & to terminate and output the optimal outcomes      & Maximum generations    \\
\bottomrule
\end{tabular}
}
\end{table*}

\subsection{Training and Test}
\label{subsec:training_and_test}

After feature selection, 3 subsets corresponding to the 3 used datasets in this study have been generated with optimal features. In this step, it's necessary to train the model with processed data so as to predict dropout.

To begin with, Long short-term memory (LSTM) is used for the training model in this study, LSTM is a special type of recurrent neural networks (RNNs). Unlike traditional feedforward neural networks and traditional RNNs, it has feedback connection and gate schema, which power LSTM to learn and memorize the information over time sequences. Likewise, human learning activities are matchable to LSTM based on gate schema. In this study, it also simulates the process of influence of past exam to the current performance. Furthermore, LSTM is capable to handle the vanishing gradient. There are multiple studies in the past focusing on Deep Knowledge Tracing to predict student performance in the quiz\citep{Chris2015} and simulation of Human Activity Recognition based on the combination of LSTM and convolutional neural network (CNN)\citep{Xia2020}.

\begin{figure}
  \centering
  \includegraphics[width=0.6\textwidth]{{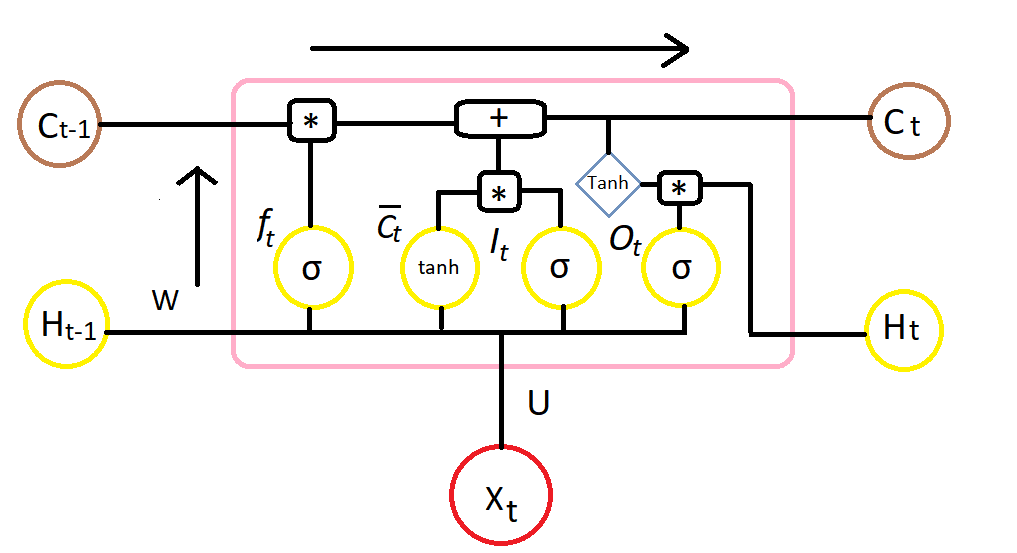}}
  \caption{The internal architecture of LSTM cell \citep{Madhu2018}}
  \label{fig:lstm}
\end{figure}

As illustrated in Figure \ref{fig:lstm}, LSTM consists of the following components which has been listed in Table \ref{tab:lstm}. The inputs at each time sequence are comprised of 3 elements, that is, $\mathit{X}_{t}$, $\mathit{H}_{t-1}$, and $\mathit{C}_{t-1}$. With regards to outputs, $\mathit{H}_{t}$ and $\mathit{C}_{t}$ are exported by the LSTM. Note that there are 8 weight $\emph{W}$ parameters in total, 4 associated with hidden state and others linked with the input. Moreover, 4 bias $\textit{b}$ will be used in one unit. All the $\emph{W}$ and $\textit{b}$ will be initialized randomly in the first place, and will be adjusted by back-propagation. To prevent gradients exploring when gradients reach the designed threshold, which is 1.01, clipping the gradients.

\begin{table*}
  \centering

  \caption{The components of a LSTM cell}
  
  \label{tab:lstm}
  \resizebox{0.7\textwidth}{!}{
\begin{tabular}{BCc^c^c}
\toprule
\rowstyle{\bfseries}
Component       & Description                       & Denote      \\
\midrule 
Forget Gate     & activation vector with sigmoid     & $\mathit{f}$   \\
Candidate layer & activation vector with Tanh       & $\mathit{\widetilde{C}}$    \\
Input Gate      & activation vector with sigmoid     & $\mathit{I}$  \\
Output Gate     & activation vector with sigmoid     & $\mathit{O}$\\
Hidden state    & hidden state vector                            & $\mathit{H}$  \\ 
Memory state    & cell state vector                  & $\mathit{C}$  \\
Current Input   & vector                            & $\mathit{X}$  \\
\bottomrule
\end{tabular}
}
\end{table*}

Once execution in LSTM during one sequence $\mathit{t}$ with using the dataset from section \ref{subsec:feature_selection}. The first move is to determine which information will be forgotten as formula \ref{for: forget} illustrates. This step is executed by the forget gate, $\mathit{H}_{t-1}$ and $\mathit{X}_{t}$ pass the gate to get a output between 0 and 1.

\begin{equation}
\mathit{f_{t}}=\sigma (\emph{W}_{\mathit{f}}\cdot \left [ \textit{H}_{t-1},\mathit{x_{t}} \right ]+\textit{b}_{f})
\label{for: forget}
\end{equation}

where $\mathit{t}$ is between 0 to 31 inclusive which indicates there are 32-time steps per student representing the information across 8 semesters and 4 courses each semester. Regarding the students who dropped out midway or have not finished all the semesters, padding time series sequences to make sure the total time step is the same. The $\mathit{x_{t}}$ is the input, whose size initial input ($\mathit{x_{0}}$) equals the number of features in the input dataset. Furthermore, there are two hidden layers and the hidden size is 50 which determines the size of the $\textit{H}_{0}$ and $\textit{C}_{0}$. 

The second step is to decide what new messages will be used for storing in the cell state. First, input gate will get the update values $\mathit{i_{t}}$, followed by a $tanh$ layer creates a vector of new candidate values $\mathit{\widetilde{C}_{t}}$, these values will be combined and the state updated. The formula for this is shown in \ref{for: input} and \ref{for: tanh}:

\begin{equation}
\mathit{i_{t}}=\sigma (\emph{W}_{\mathit{i}}\cdot \left [ \textit{H}_{t-1},\mathit{x_{t}} \right ]+\textit{b}_{i})
\label{for: input}
\end{equation}
\begin{equation}
\mathit{\widetilde{C}_{t}}=tanh (\emph{W}_{\mathit{C}}\cdot \left [ \textit{H}_{t-1},\mathit{x_{t}} \right ]+\textit{b}_{C})
\label{for: tanh}
\end{equation}

Next step to update the $\mathit{C}_{t-1}$, and create a new cell state $\mathit{C}_{t}$. According to the formula \ref{for: cell_State}, combining the previous outputs $\mathit{f_{t}}$, $\mathit{i_{t}}$, and $\mathit{\widetilde{C}_{t}}$, by a series of arithmetic operations, to get the final cell state $\mathit{C}_{t}$, which will be the input in the next step.

\begin{equation}
\mathit{C}_{t}=\mathit{f}_{t}*\mathit{C}_{t-1}+\mathit{I}_{t}*\widetilde{C}_{t}
\label{for: cell_State}
\end{equation}

The final step is to output the hidden state. As shown in formula \ref{for: output} and \ref{for: hidden}, similar to step 1 and 2, carrying $\mathit{H}_{t-1}$ and $\mathit{X}_{t}$ as inputs, to pass output gate with sigmoid which will make a decision regarding the parts of outputs. Then output is transited to a $tanh$ layer to acquire the hidden state $\mathit{H}_{t}$, which will be used as the input in the time step or another neural network.

\begin{equation}
\mathit{O_{t}}=\sigma  (\emph{W}_{\mathit{o}}\cdot \left [ \textit{H}_{t-1},\mathit{x_{t}} \right ]+\textit{b}_{o})
\label{for: output}
\end{equation}

\begin{equation}
\textit{H}_{t}=\mathit{O}_{t}*tanh(\mathit{C}_{t})
\label{for: hidden}
\end{equation}

In this study, after a series of hyper-parameter tuning, the final output hidden state will be regarded as the input of a 3-layer fully connected (FC) neural network to predict the dropout. The hidden size in FC is 128, in the first layer, the sigmoid is deployed. Likewise, ReLU is applied in the output layer. As for loss measurement, as illustrated in formula \ref{for: MSE}, Mean Squared Error (MSE) is used to compute the loss:

\begin{equation}
\mathit{MSE}=\frac{1}{\mathit{n}}\sum_{\mathit{i}=1}^{\mathit{n}}(\mathit{Y}_{\mathit{i}}-\hat{\mathit{Y}_{\mathit{i}}})^{2}
\label{for: MSE}
\end{equation}

where $\mathit{n}$ is the number of samples, $\mathit{Y}_{\mathit{i}}$ is the observed value, and $\hat{\mathit{Y}_{\mathit{i}}}$ is the predicted value.

\begin{figure}
  \centering
  \includegraphics[width=0.4\textwidth]{{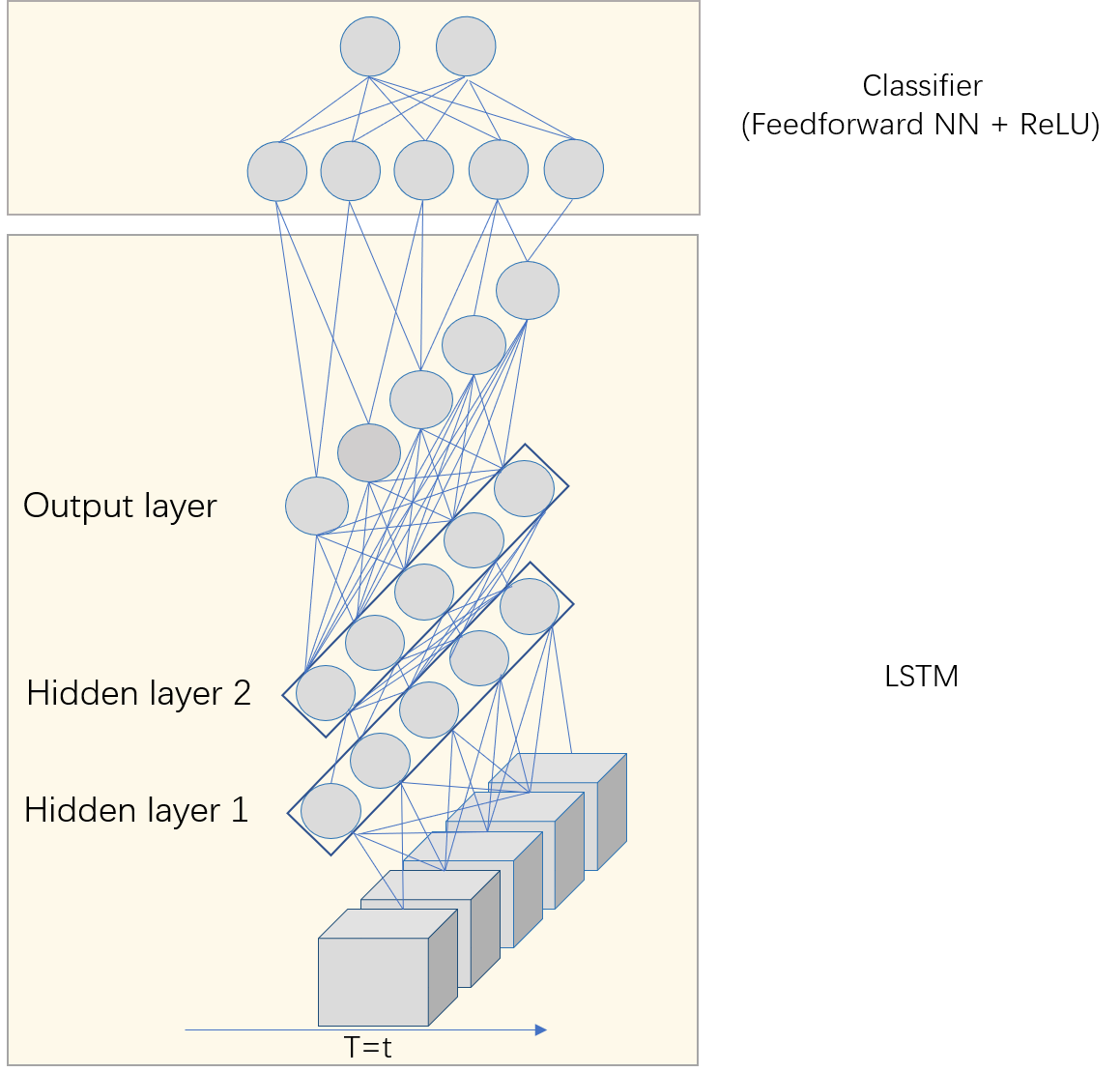}}
  \caption{The workflow of LSTM + FCs}
  \label{fig:lstm+snn}
\end{figure}

After computing the loss, the back-propagation will be executed to adjust the weights by using Adam as the optimizer. Furthermore, by introducing a dropout layer on outputs of each LSTM layer apart from the last layer, with a dropout possibility equal to 0.7, to prevent overfitting. Figure \ref{fig:lstm+snn} visualizes the process of once training.

\section{Curriculum Semantic Analysis}
\label{sec:Curriculum_semantic_Analysis}
Identifying dropout is an important step to understand the reason why students fail and what can be done to increase graduation rates. We argue that the sequence of courses taken by students may influence student attrition. This section presents an approach to create a sequence of courses based on its descriptions. The idea is that in the future be able to correlate dropout rates, the sequence of courses taken by students and the course content/description.

We divide this section into two parts. The first part introduces methods to measure the similarity between course pairs using the Bidirectional Encoder Representations from Transformers (the so-called BERT language model). The second part is responsible for ordering course pairs by employing Semi-Reference Distance (SemRefD) \citep{Manrique2019}. 

\subsection{Procedure Overview}
\label{subsec:procedure_overview}
\begin{figure}
  \centering
  \includegraphics[width=0.8\textwidth]{{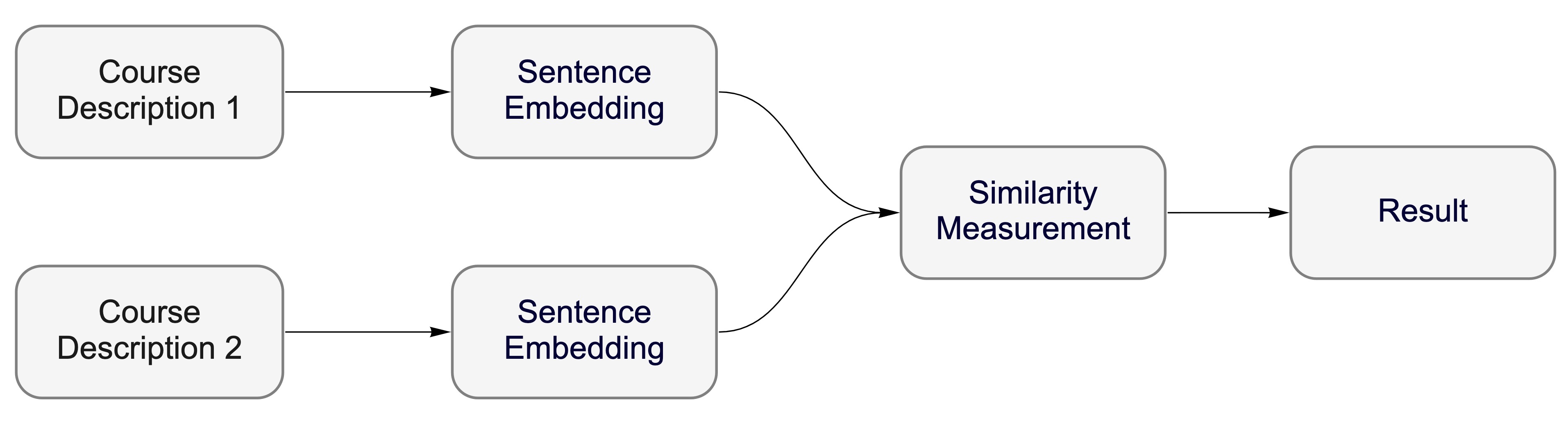}}
  \caption{The brief workflow of the Similarity Measurement}
  \label{fig:bert}
\end{figure}

\begin{figure}
  \centering
  \includegraphics[width=0.8\textwidth]{{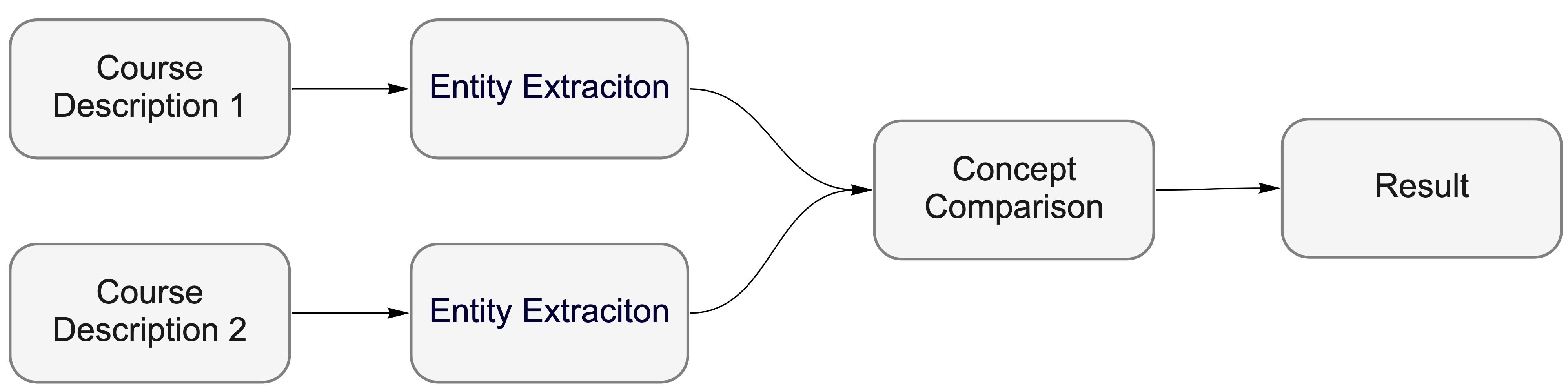}}
  \caption{Prerequisite Identification workflow}
  \label{fig:semrefd}
\end{figure}

As Figure \ref{fig:bert} and \ref{fig:semrefd} show, the first move is to unify BERT to conduct similarity measurement. It involves three steps, that is, dataset acquirement, sentence embedding, and similarity measurement. Likewise, the second step uses the same dataset, the first process is to extract entities, followed by concept comparison and the sequence will be identified between courses in the end.

\subsection{Similarity Measurement}
\label{subsec:similarity_measurement}

To compare the courses description, it is necessary to encode the contextual data into vectors so as to be comparable in a semantic way. To achieve this, we make use of BERT for sentence embedding. 

BERT was proposed by Google \citep{Jacob2018}, which is a revolutionary pre-training model that uses multi-head attention-based transformer to learn contextual relations between words or sentences in the context. Transformers were also proposed by Google \citep{Ashish2017}. It consists of an encoder, which is a bidirectional RNN, and a decoder. There is a scaled dot-product attention layer and a feedforward neural network layer in the encoder. With regards to self-attention, it unifies the matrix representation to calculate scoring and final output embedding in one step as formula \ref{for: self-attention} illustrates:

\begin{equation}
Attention(Q, K, V) = softmax(\frac{QK^T}{\sqrt{d_k}})V
\label{for: self-attention}
\end{equation}

where $K$ and $V$ create a key-value pair of input of dimension $d_k$, $Q$ stands for the query. The output is a weighted sum of values.

Compare with computing the attention one time, the multi-head mechanism which is used in BERT, goes through the scaled dot-product attention several times at the same time and creates separate embedding matrices that are combined into one final output embedding as formula \ref{for: multi-head-attention} shown:

\begin{equation}
MultiHead(Q, K, V) = Concat(head_1, ..., head_h)W^O
\label{for: multi-head-attention}
\end{equation}

where $head_i = Attention(Q W^Q_i, K W^K_i, V W^V_i)$ and $W^Q_i$, $W^K_i$, $W^V_i$, and $W^O$ are separate weight matrices.

\begin{figure}
  \centering
  \includegraphics[width=0.6\textwidth]{{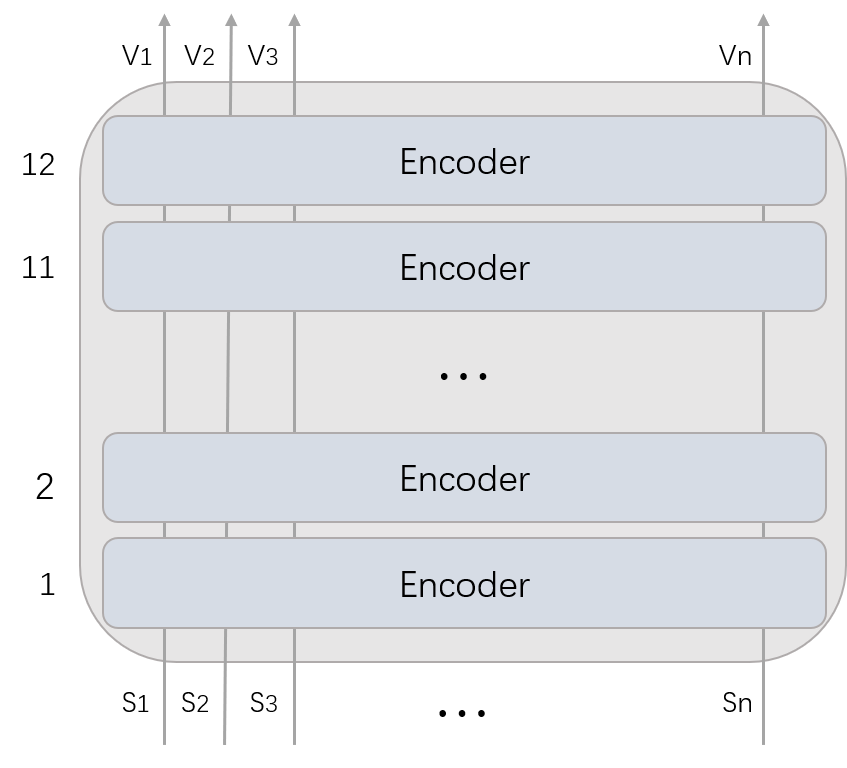}}
  \caption{BERT Base}
  \label{fig:BERT1}
\end{figure}

BERT contains 12 stacked encoders in the base version, 24 stacked encoders in a larger version. Transformer encoders read each sequence of words at once instead of the sequential reading of text input as in directional models. With this characteristic, a model is able to determine a word's context based on all of its surrounding information as shown in Figure \ref{fig:BERT1}.

To start with, the datasets this study uses ANU Course \& Program Website\footnote{\url{https://programsandcourses.anu.edu.au/}}. Furthermore, this study takes Computer Science courses into consideration, hence, course code beginning with "COMP" will be considered. After using a tokenizer to segment the full text into sentences and applying BERT on sentences as such for encoding, the vectors of sentences are constructed. Next, the vectors will be computed by cosine similarity measurement which aims to get the distance, the opposite of similarity as equation \ref{for: cosine-similarity} specifies:

\begin{equation}
\cos ({\bf t},{\bf e})= {{\bf t} {\bf e} \over \|{\bf t}\| \|{\bf e}\|} = \frac{ \sum_{i=1}^{n}{{\bf t}_i{\bf e}_i} }{ \sqrt{\sum_{i=1}^{n}{({\bf t}_i)^2}} \sqrt{\sum_{i=1}^{n}{({\bf e}_i)^2}} }
\label{for: cosine-similarity}
\end{equation}

where $\bf t$ and ${\bf e}$ are sentence vectors which were generated by BERT, the output ranges within [-1, 1] depicting the degree of contextual similarity.

To get the entire contextual similarity, the average similarity among sentences will be calculated. After the processes above, the course similarity is obtained.

\subsection{Prerequisite Identification}
\label{subsec:prerequisite_identification}

Despite the similarity between courses having been computed, the sequence between highly similar courses remains uncertain, hence, identifying the prerequisite dependency (PD) between two highly similar courses becomes necessary. 

Regarding PD, it is a relation between two concepts where, in education, the prerequisite concept should be taught first. For instance, \textit{Binary Tree} and \textit{Red-black Tree} are two concepts belonging to the data structure field in Computer Science, the latter should be introduced after the former. By measuring the prerequisite relationship between courses, the curriculum will be analyzed as a whole.

To begin with, similar to the similarity measurement, the same dataset, and pre-processing techniques were used in this study. Subsequently, the entities behind the text will be extracted, and the tool named \textit{TextRazor} was employed to complete this task. \textit{TextRazor} is an NLP-based API developed to segment text and capture conceptual terms. Then a technique called Semi-Reference Distance (SemRefD) was conducted to measure semantic PD between the entities of two courses in DBpedia\footnote{\url{https://www.dbpedia.org/}} Knowledge Graph (KG).

With respect to KG, it is known as a semantic network, which stands for a real-world entities network, in this study, they refer to concepts, and illustrate the relationship between them. DBpedia is one of the main KGs on the Semantic Web and provides a wide variety of topics that can be used to encompass courses from various fields of study. Moreover, it is also very tolerant and inclusive of many different semantic properties, which empower the liberal connection to multiple types of concepts. Using the given concept as a query in DBpedia, there will be two lists storing the candidate concepts, that is, the direct list and the neighbor list. As for a direct list, in which a list of concepts sharing a category with the given concept will be returned. For the latter list, in which the candidate list is expanded by adding concepts linked to the target through non-hierarchical pathways up to $m$ hops \citep{Ruben2019}. The path length parameter $m$ decides the maximum length of the path between the target concept and the farthest candidate concept to consider, in this study, $m$ is 1. Soon after acquiring the candidate lists, the SemRefD will be performed to compute the degree of prerequisite in the next step.

SemRefD was presented by Manrique et al.\citep{Ruben2019} based on Reference Distance (RefD) which was proposed by Chen et al. \citep{Liang2018} as defined in formula \ref{for: RefD} by inputting two concepts which are denoted $c_A$ and $c_B$ as below:

\begin{equation}
RefD(c_A,c_B)=\frac{\sum_{j=1}^{k}i(c_j,c_B)s(c_j,c_A)}{\sum_{j=1}^{k}s(c_j,c_B)}-\frac{\sum_{j=1}^{k}i(c_j,c_A)s(c_j,c_B)}{\sum_{j=1}^{k}s(c_j,c_B)}
\label{for: RefD}
\end{equation}

\begin{figure}
  \centering
  \includegraphics[width=0.4\textwidth]{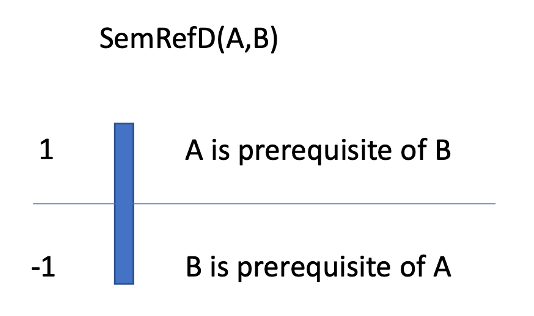}
  \caption{The prerequisite of concept A ($c_A$) and Concept B ($c_B$)}
  \label{fig:prerequisite}
\end{figure}

An indicator function $i(c_j,c_A)$ indicates that there is a relationship between $c_j$ and $c_A$, and a weighting function $s(c_j,c_A)$ indicates whether there is a relationship. The values of $RefD(c_A,c_B)$ range from -1 to 1. According to Figure \ref{fig:prerequisite}, $c_B$ is more likely to be a prerequisite for $c_A$ the if it is closer to 1.

RefD does not take into account the semantic properties of DBpedia to determine whether two concepts have a prerequisite dependency. SemRefD does. In the weighting function $1(c_j,c_A)$, the common neighbors' concepts in the KG KG hierarchy are considered, while in the indicator function $1(c_j,c_A)$, the property paths between concept target and related concepts are considered \citep{Ruben2019}.

As a result, all concepts from the two courses will be compared and summed up to reveal their order of them, that is, whether A is a prerequisite of B or B is a prerequisite of A.


\chapter{Results and Discussion}
\label{cha:result}
\setcounter{tocdepth}{3}
\setcounter{secnumdepth}{3}
In this chapter, the results will be presented in 2 sections from experiments described in Chapter \ref{cha:methodology}, namely, dropout prediction and curriculum semantic analysis along with discussions in detail. Notably, the results yielded by methods in the systematic review have been indicated in Section \ref{sec:systematic_review}. Additionally, the experimental environment will be introduced initially.

\section{Experimental Environment}
\label{sec:experimental_env}
The experiments are conducted on the following devices and corresponding hardware as shown in Table \ref{tab:hardware}. Specifically, the paper retrieving process in systematic review is completed on the Apple Macbook Pro. The rest of the experiments are conducted on the server.

\begin{table*}
  \centering

  \caption{Experimental Environment}
  
  \label{tab:hardware}
  \resizebox{\textwidth}{!}{
\begin{tabular}{BCc^c^c^c^c^c}
\toprule
\rowstyle{\bfseries}
Model       & CPU      & GPU & Memory & Hard disk size  & System    \\
\midrule 
Server     & Intel(R) Xeon(R) CPU E5-2698 v4 @ 2.20GHz     & Tesla V100-DGXS-32GB*4 & 251GB & 10TB & Ubuntu 18.04.6  \\
Apple Macbook Pro & M1 chip  & 8-core GPU & 16GB & 512GB & macOS Monterey 12.0.1   \\
\bottomrule
\end{tabular}
}
\end{table*}

\section{Dropout Prediction}
\label{sec:dropout_prediction}

This section contains two subsections. The evaluation criteria will be presented firstly, followed by the results from feature selection and dropout prediction.

\subsection{Evaluation  Metrics}
\label{subsec:direct_cost}
\begin{figure}
  \centering
  \includegraphics[width=0.8\textwidth]{{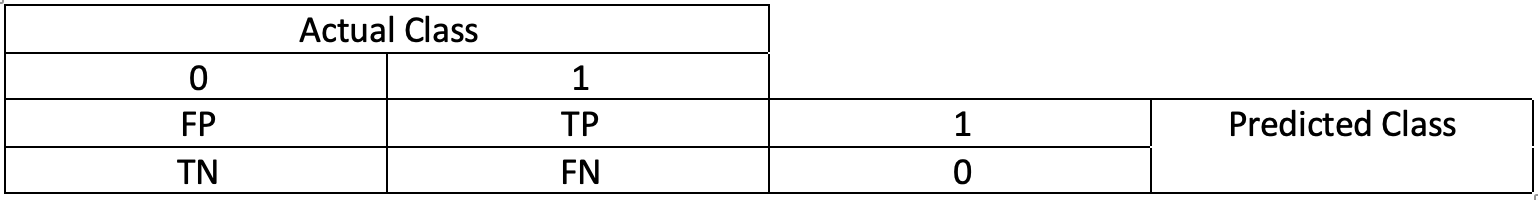}}
  \caption{Confusion Matrix}
  \label{fig:matrix}
\end{figure}

Dropout prediction is evaluated by accuracy, precision, recall, and F1 score based on the confusion matrix as shown in Figure \ref{fig:matrix}. Specifically, the $TP$, $TN$, $FP$, and $FN$ are defined as below:

True Positive ($TP$):  Models predicted positive outcomes and are accurate.

True Negative ($TN$): Models predicted negative outcomes and are accurate.

False Positive ($FP$): Models predicted positive outcomes and are inaccurate.

False Negative ($FN$): Models predicted negative outcomes and are inaccurate.

To be specific, accuracy refers to the proportion of correctly predicted observations in whole samples. The precision measures the correctly predicted TP samples to the total positive ones. Recall depicts the sensitivity of the model by measuring the correctly predicted positive observations to all the observations. F1 score measure the overall performance of model The equations are as formula \ref{for: acc} - \ref{for: f1} shown below:

\begin{equation}
Accuracy=\frac{TP+TN}{TP+FP+FN+TN}
\label{for: acc}
\end{equation}
\begin{equation}
Precision=\frac{TP}{TP+FP}
\label{for: precision}
\end{equation}
\begin{equation}
Recall=\frac{TP}{TP+FN}
\label{for: recall}
\end{equation}
\begin{equation}
F1=\frac{2*(Recall*Precision)}{Recall + precision}=\frac{TP}{TP+\frac{FP+FN}{2}}
\label{for: f1}
\end{equation}

\subsection{Experimental Results an Discussions}
\label{subsec:result}
As mentioned in Section \ref{subsec:data_and_pre-processing}, ADM, ARQ, and CSI datasets will be used to evaluate the proposed models in terms of feature selection and dropout prediction. The first section is regarding feature selection, which was proposed in Section \ref{subsec:feature_selection}, followed by dropout prediction, as mentioned in section \ref{subsec:training_and_test}.

\subsubsection{Feature Selection}
\label{subsubsec:featrure_selection_res}
The results of feature selection as listed in Table \ref{tab:feature_selection}. After 100 generations of evolution in a population comprised of 1000 individuals, as for the performance of final optimal individuals among these datasets, ARQ has the best performance in terms of accuracy (90.61), F1 score (95.03) with the least number of dropped features, indicating that this individual fits the environment most than the others. By comparison, CSI has the least performance in terms of measurement metrics. Observing the lower score in precision than recall across used datasets, indicates that the model has its bias in terms of the prediction preference in positive instead of negative, and emphasises the importance of the balance of datasets. Furthermore, by acquiring the ratio of negative in CSI (24\%), ADM (13\%), and ARQ (10\%), also proves this inference. Thus, employing a model on a balanced dataset may improve its performance. As for ADM, it has the nearly same recall as ARQ's and the best recall score in this experiment. Overall, the results turn out this model align with this experiment.

\begin{table*}
  \centering

  \caption{Results of ADM, ARQ, and CSI by SVM}
  
  \label{tab:feature_selection}
  \resizebox{\textwidth}{!}{
\begin{tabular}{BCc^c^c^c^c^c}
\toprule
\rowstyle{\bfseries}
Dataset     & Accuracy(\%) & Precision(\%) & Recall(\%) & F1 score(\%) & No. dropped features         \\
\midrule 
ADM      &88.42  &88.75 &\textbf{99.52} &93.83 & 10           \\
ARQ    & \textbf{90.61}  &\textbf{91.14} &99.26 &\textbf{95.03} & 6          \\
CSI     &81.23  &82.28 &97.84 &89.36 & 12       \\
\bottomrule
\end{tabular}
}
\end{table*}

\subsubsection{Dropout Prediction}
\label{subsubsec:dropout_prediction_res}
\begin{figure}[H]
\centering
\begin{minipage}{.5\textwidth}
  \centering
  \includegraphics[width=\textwidth]{{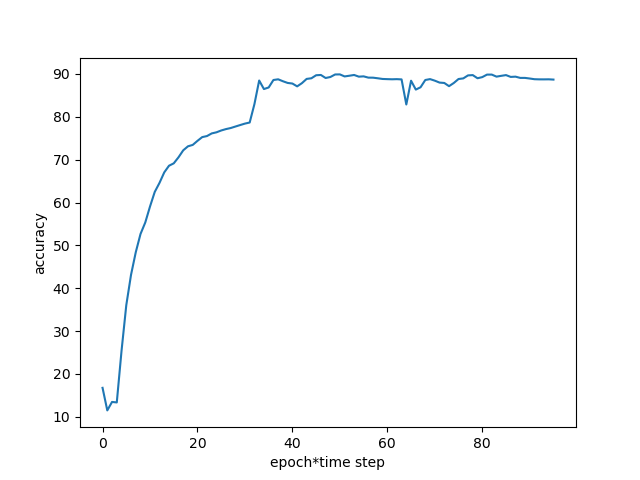}}
  \caption{Accuracy of LSTM+FC on ADM}
  \label{fig:ADM_acc}
\end{minipage}%
\begin{minipage}{.5\textwidth}
  \centering
  \includegraphics[width=\textwidth]{{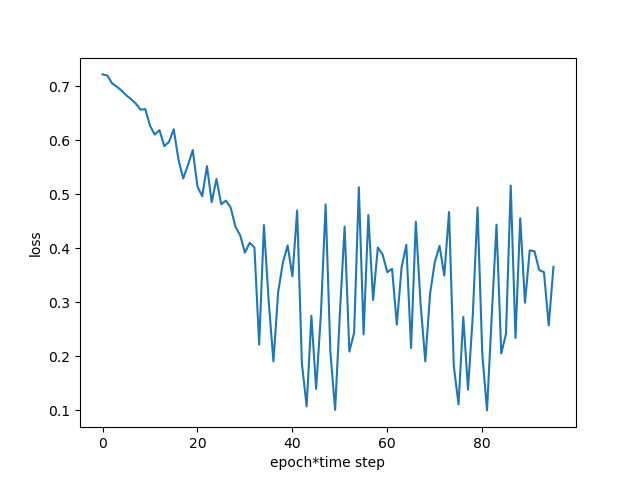}}
  \caption{Loss of LSTM+FC on ADM}
  \label{fig:ADM_loss}
\end{minipage}
\end{figure}

\begin{figure}[H]
\centering
\begin{minipage}{.5\textwidth}
  \centering
  \includegraphics[width=\textwidth]{{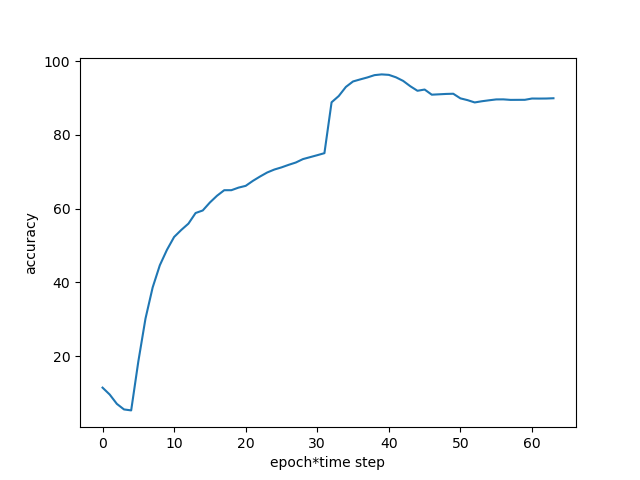}}
  \caption{Accuracy of LSTM+FC on ARQ}
  \label{fig:ARQ_acc}
\end{minipage}%
\begin{minipage}{.5\textwidth}
  \centering
  \includegraphics[width=\textwidth]{{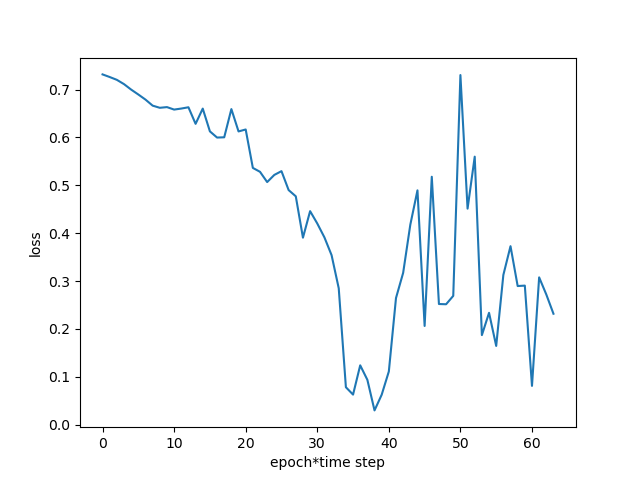}}
  \caption{Loss of LSTM+FC on ARQ}
  \label{fig:ARQ_loss}
\end{minipage}
\end{figure}

\begin{figure}[H]
\centering
\begin{minipage}{.5\textwidth}
  \centering
  \includegraphics[width=\textwidth]{{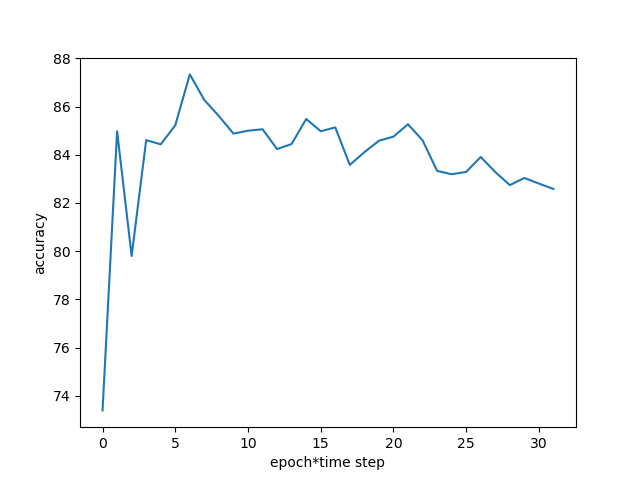}}
  \caption{Accuracy of LSTM+FC on CSI}
  \label{fig:CSI_acc}
\end{minipage}%
\begin{minipage}{.5\textwidth}
  \centering
  \includegraphics[width=\textwidth]{{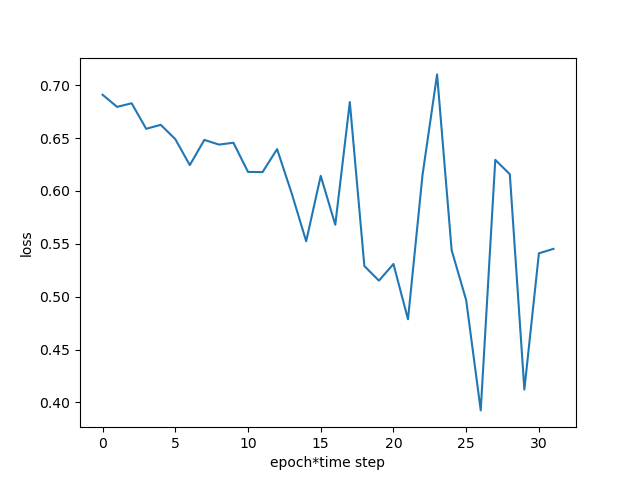}}
  \caption{Loss of LSTM+FC on CSI}
  \label{fig:CSI_loss}
\end{minipage}
\end{figure}

Likewise, this experiment used the same datasets on the dropout prediction. As dropout prediction is one of the main objectives of this project, the performance of the proposed model is important to this project. As illustrated in \ref{fig:ADM_acc} - \ref{fig:CSI_loss}, these outcomes on training reveal the capacity of the model, whose end of the abscissa is the $epoch(every 10 epochs) * time step$. Furthermore, the performance of the model on test further validates the suitability of the model on datasets as shown in Table \ref{tab:lstm_res}.

According to the accuracy and loss during training, we can observe that the model always converges after 10 epochs. As for the accuracy in ADM and ARQ datasets, which starts from below 10\% initially, then fall to the lower, and rises to close to the top afterwards. For ARQ, the curve is similar to some extent. After investigation, multiple reasons are identified, which leads to the abnormal curve. For instance, the dropout rate is high, which shows that the model drops the well-functional neurons, or the batch size is large. In this study, the time step is fixed as mentioned in section \ref{subsec:training_and_test}, which indicates that the batch size is uncontrollable. In addition, after hyper-parameter tuning, high dropout rate (0.7) enables the best results of the model, which shows that selecting the current dropout rate is a trade-off decision. Thus, this study chose to keep a better performance by using the current dropout rate. 

In terms of loss, after reaching the convergence, it will repeat the loss curve along with the rest of the time step. Apart from the repetition of loss, we also observed the unstable curve during one epoch training, which is named multiple descent \citep{Lin2020}. The reason behind the anomaly may vary. For example, we assumed there are minimas $\theta _{1}$ and $\theta _{2}$, when the distance between two minima $d=\left \| \theta _{1}-\theta _{2} \right \|$ is very small, and the learning rate is not small enough, which leads to cross a local $\theta _{1}$ and arrives in $\theta _{2}$ eventually. Also, this phenomenon can be caused by datasets \citep{Lin2020}.

With respect to the performance of the proposed model in the test, the model performs well in ADM and ARQ, whose best accuracy reaches 92.83\% and 97.65\% respectively. Notably, the accuracy of ARQ improves the result in Manrique's previous work (95.2\%) by 2.45\% \citep{Ruben2019}. Finally, we captured a potential improvement for the model by adopting dynamic time steps instead of fixed steps to make full use of the dataset. Overall, the model proposed by this study is suitable for the current datasets. 

\begin{table*}
  \centering

  \caption{Results of ADM, ARQ, and CSI }
  
  \label{tab:lstm_res}
  \resizebox{\textwidth}{!}{
\begin{tabular}{BCc^c^c^c^c}
\toprule
\rowstyle{\bfseries}
Dataset     & Avg. acc (10 iter.)(\%) & Avg. acc (100  iter.)(\%) & Avg. acc (200  iter.)(\%) & Top. acc(\%)     \\
\midrule 
ADM      & 88.76    & 88.24 & 90.17 &  92.83  \\
ARQ    & \textbf{94.84}   & \textbf{95.76} & \textbf{95.19} & \textbf{97.65}              \\
CSI    & 74.98      & 75.84 & 76.31 & 81.52 \\
\bottomrule
\end{tabular}
}
\end{table*}

\section{Curriculum Semantic Analysis}
\label{sec:curriculum_semantic_analysis}
This section contains two subsections, that is, similarity measurement, as mentioned in Section \ref{subsec:similarity_measurement}, and prerequisite identification as mentioned in Section \ref{subsec:prerequisite_identification}.

\subsection{Similarity Measurement}
\label{subsec:similarity_measurement_res}
\begin{figure}
  \centering
  \includegraphics[width=\textwidth]{{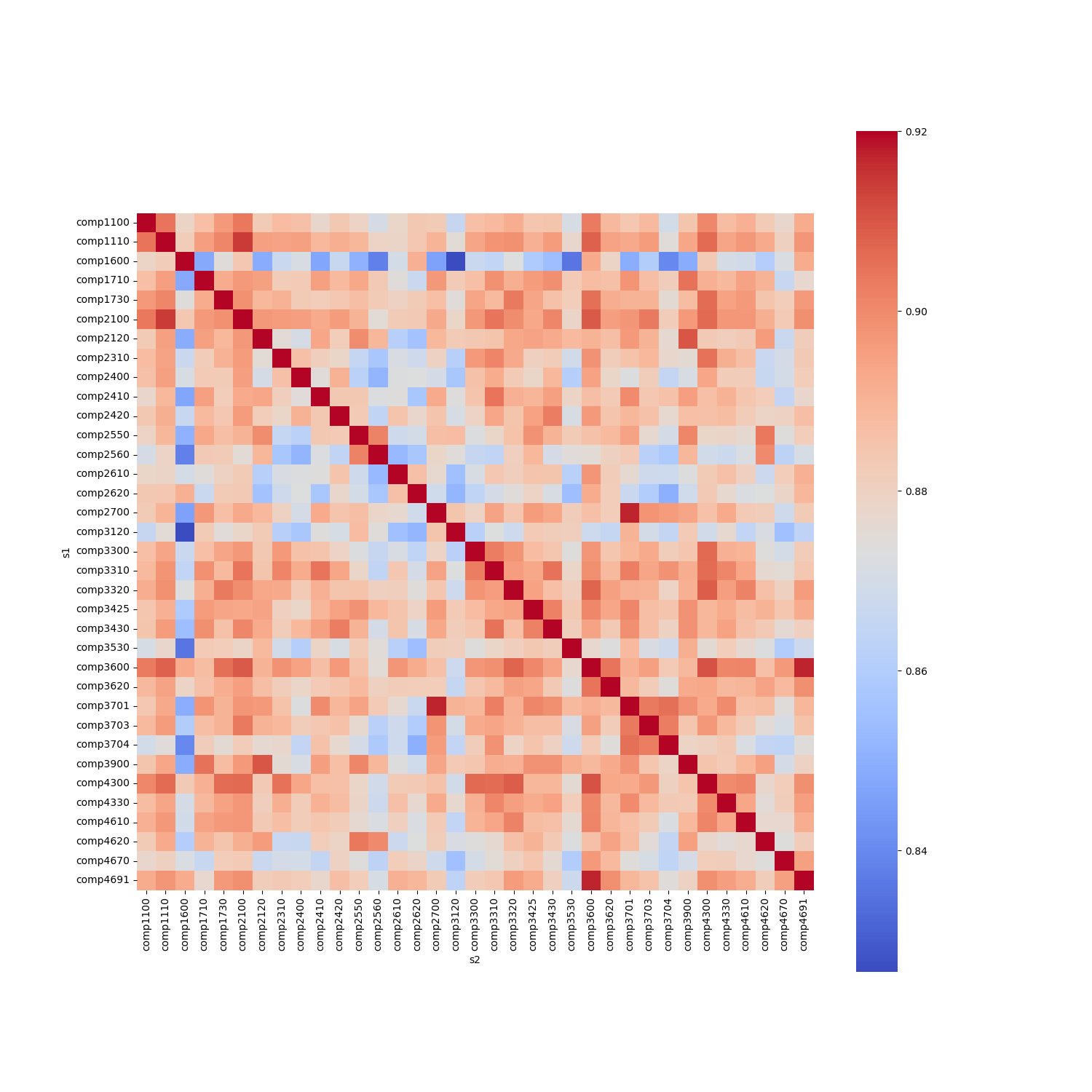}}
  \caption{All level courses similarity}
  \label{fig:Heatmap_output}
\end{figure}

\begin{figure}[H]
\centering
\begin{minipage}{.5\textwidth}
  \centering
  \includegraphics[width=\textwidth]{{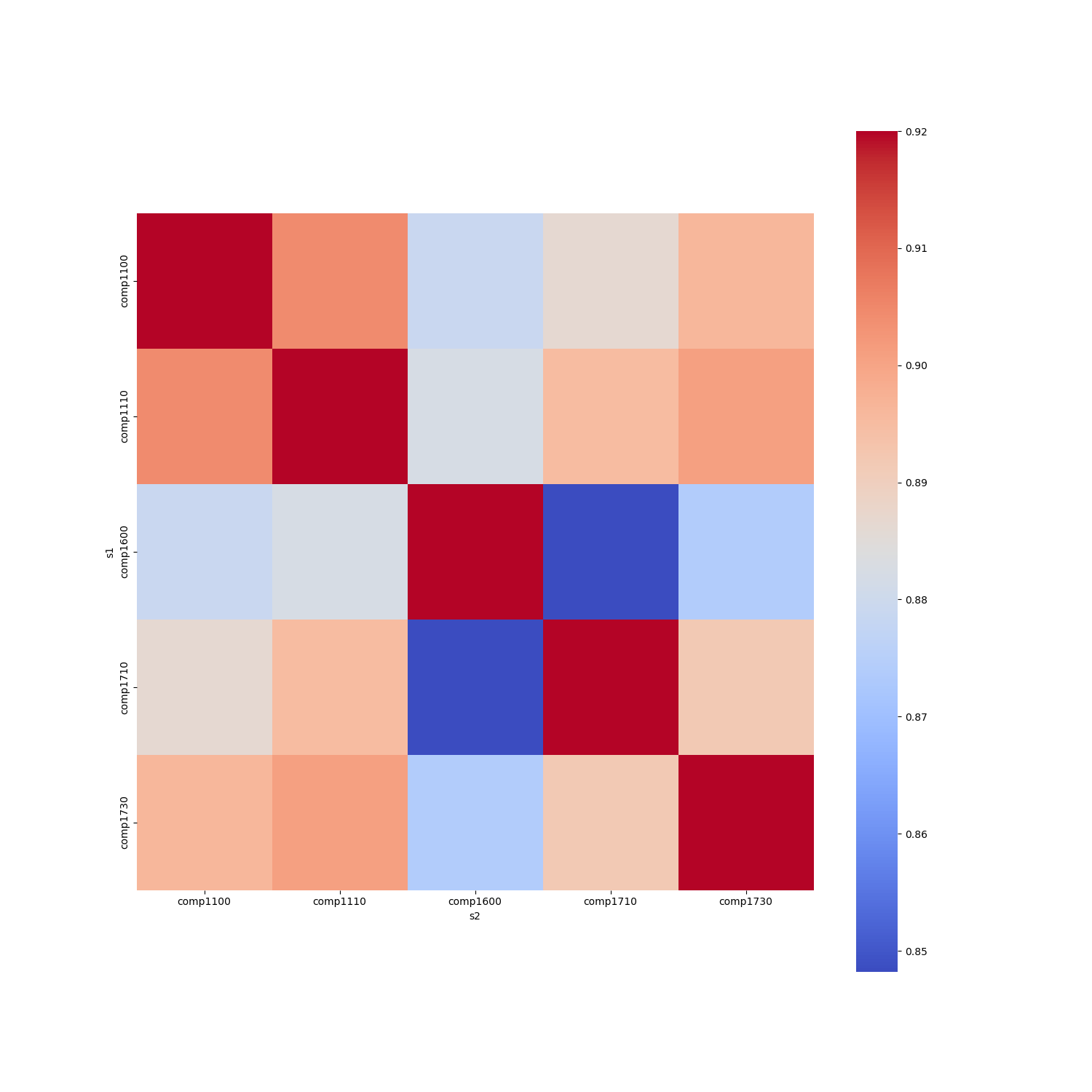}}
  \caption{1000-level courses similarity}
  \label{fig:output_1000}
\end{minipage}%
\begin{minipage}{.5\textwidth}
  \centering
  \includegraphics[width=\textwidth]{{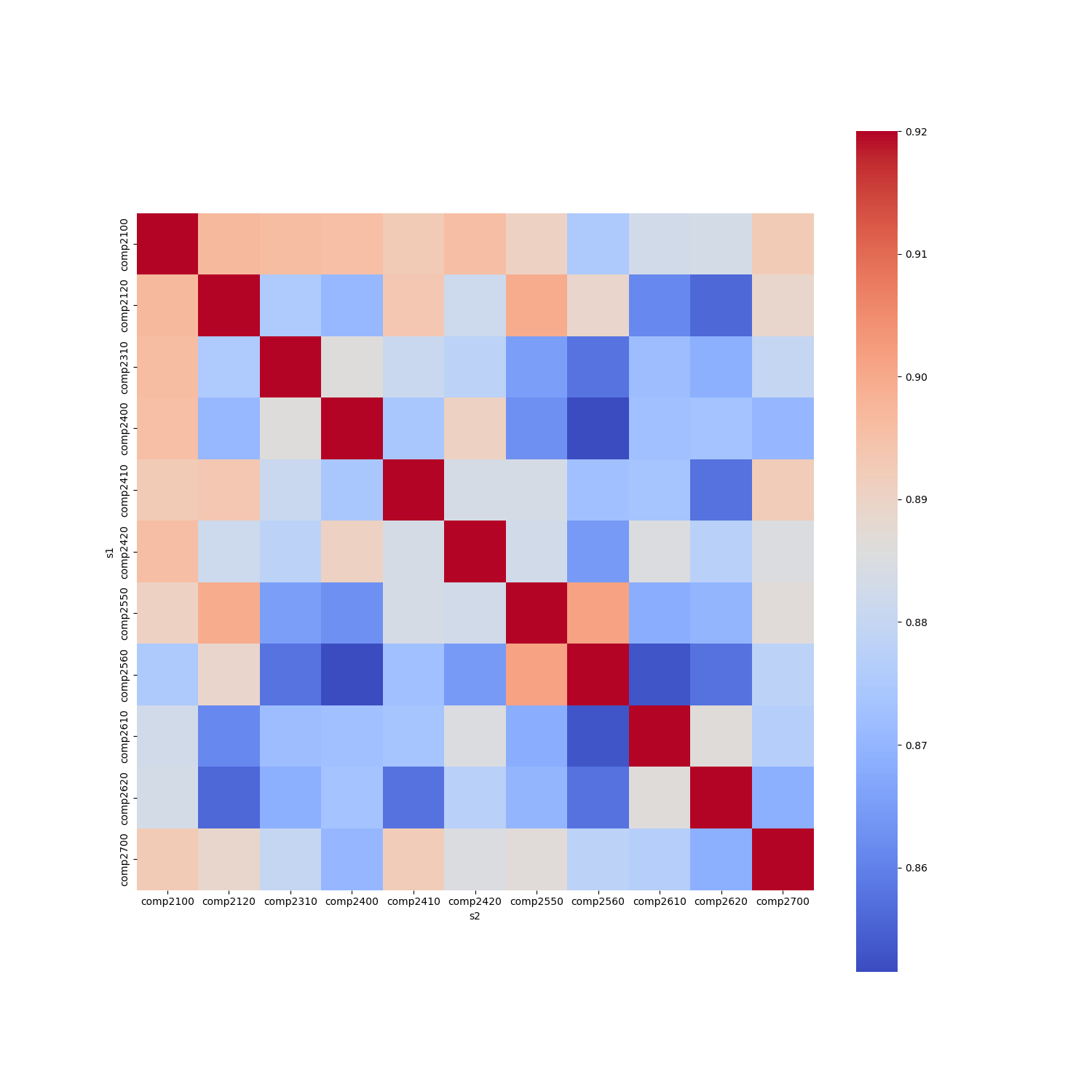}}
  \caption{2000-level courses similarity}
  \label{fig:output_2000}
\end{minipage}
\end{figure}

\begin{figure}[H]
\centering
\begin{minipage}{.5\textwidth}
  \centering
  \includegraphics[width=\textwidth]{{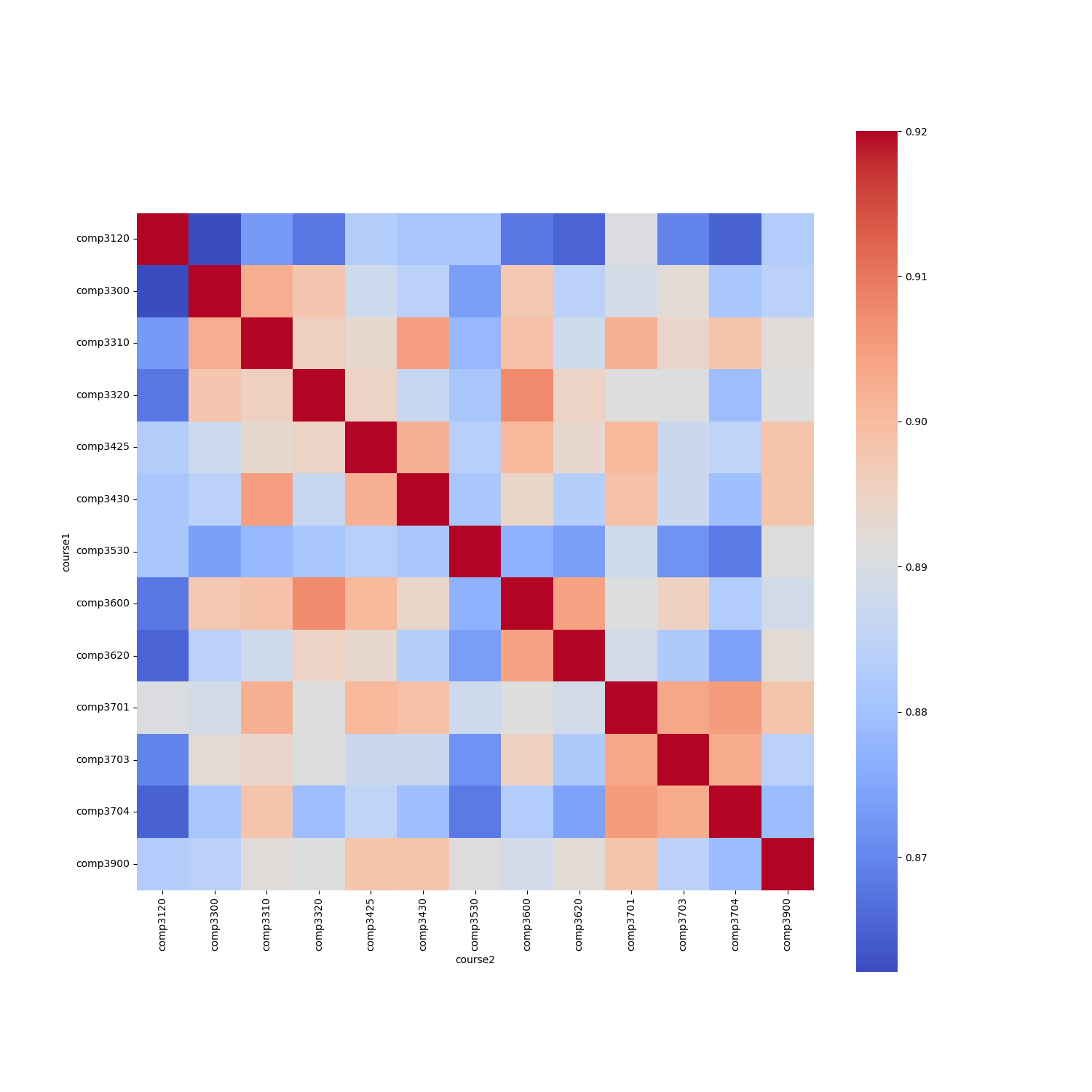}}
  \caption{3000-level courses similarity}
  \label{fig:output_3000}
\end{minipage}%
\begin{minipage}{.5\textwidth}
  \centering
  \includegraphics[width=\textwidth]{{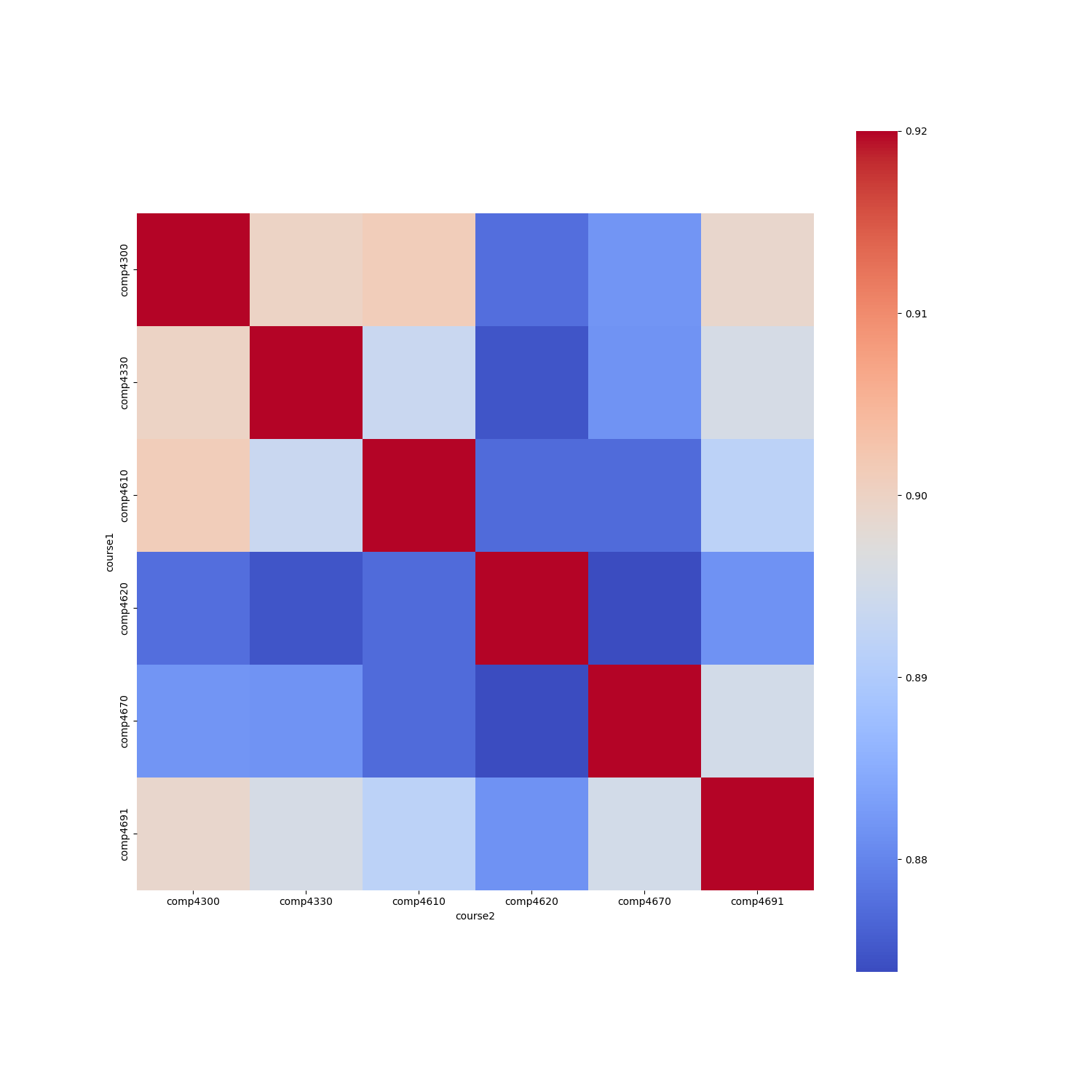}}
  \caption{4000-level courses similarity}
  \label{fig:output_4000}
\end{minipage}
\end{figure}

After the encoding between sentences captured by BERT and the average similarity between courses computed hereby, the results have been acquired and visualized in heat map format from Figure \ref{fig:Heatmap_output} ranging from 0.8265 to 1.0, which lists the result of all the course comparison, to Figure \ref{fig:output_1000} - Figure \ref{fig:output_4000} as below.

In 1000-level courses comparison, we can see that COMP1110 (Structured Programming) has the closest average distance to the rest of courses. In contrast, the COMP1600 (Foundations of Computing) has the farthest from the others. From course content to interpretation, similarity among COMP1100 (Programming as Problem Solving) to COMP1100 is one of the most fundamental programming courses among 1000-level courses, and COMP 1600 focuses on the mathematical perspective.

With respect to the 2000-level course comparison, COMP2100 (Software Design Methodologies) has the closest average distance to the rest of the courses. On the contrary, the COMP2560 (Studies in Advanced Computing R \& D) has the farthest to others. To sum up, the overall similarity between 2000-level courses becomes lower than 1000-level's, which indicates that curriculum differentiation has appeared.

As regards 3000-level and 4000-level courses, it can be seen that this trend has become more obvious and grown, which also aligns with real situations, for example, students' knowledge grows, the course content deepens and divided by specifications, such as Data Science, Machine Learning, etc. In the meantime, students also have an interest to dive in, which aids students to make use of strengths in the region of interest and improve their academic performance in return. Finally, we identified an obstacle in automatic evaluation, which is a common limitation in studies in systematic review \citep{Piedra2018}, \citep{Pata2013}, \citep{Yu2007}, \citep{Saquicela2018SimilarityDA} and \citep{Tapia2018}.

\subsection{Prerequisite Identification}
\label{subsec:prerequisite_identification_res}

As Section \ref{subsec:similarity_measurement_res} presented, the similarity course has been computed, thus it is necessary to identify the prerequisite between two similar courses.

\begin{figure}
  \centering
  \includegraphics[width=0.8\textwidth]{{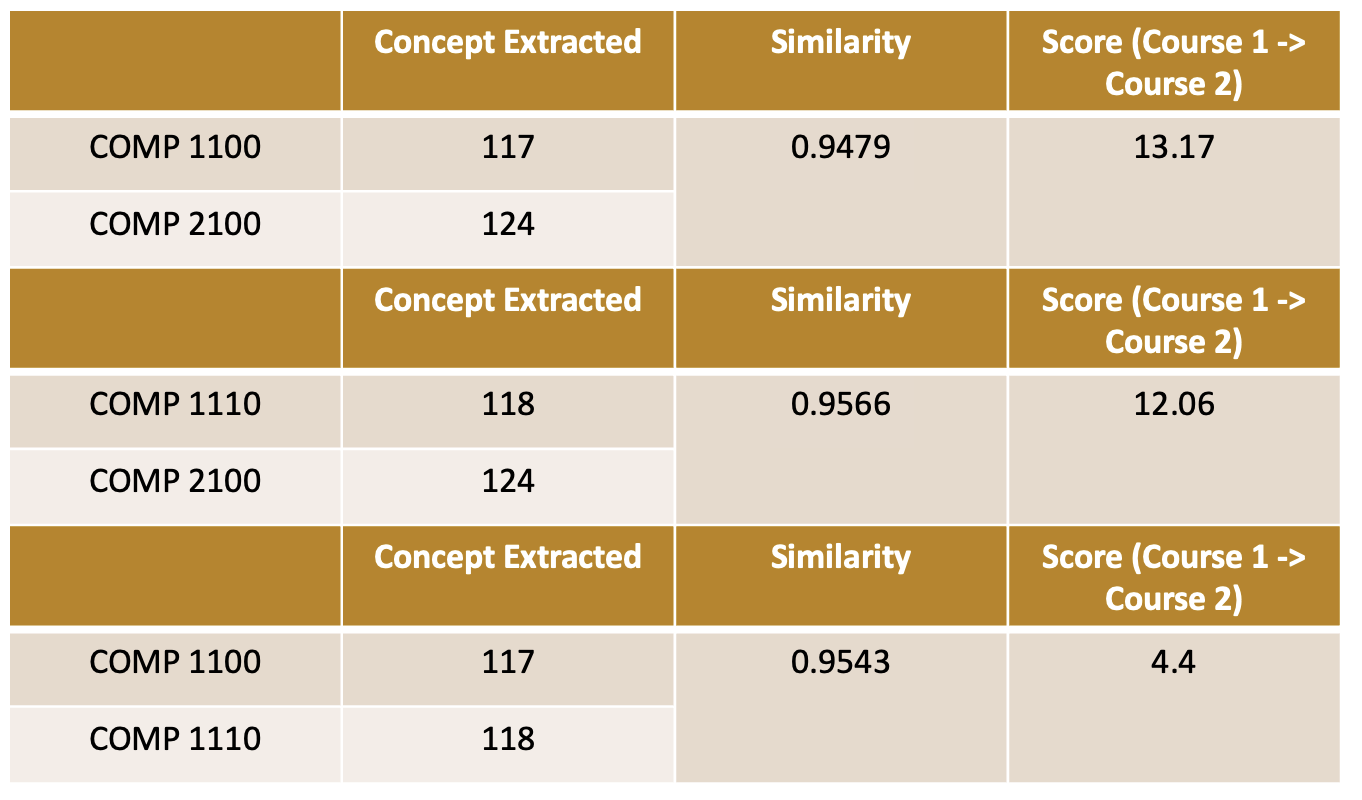}}
  \caption{All level courses similarity}
  \label{fig:pre_res}
\end{figure}

For this study, we selected 3 courses, that is, COMP1100, COMP1110, and COMP2100 to conduct this experiment as they have high similarity scores in the previous stage and they are all programming-oriented courses. The result is shown in Figure \ref{fig:pre_res}.

Between COMP1100 and COMP2100, whose similarity is 0.9479, the prerequisite score is 13.17. Based on the rule of thumb in the experiment, the score is very high, indicating that there is a strong relation in prerequisite in these two courses.

Similarly, COMP1110 and COMP2100 still have a very high score, which is 12.06, indicating that COMP1110 is also one of the prerequisites of COMP2100.

Regarding COMP1100 and COMP1110, the similarity is 0.9543 and the score is 4.4, which reveals that although there is a strong bond between these two courses, the prerequisite relationship is not as strong as the relationship with COMP2100 respectively.

\chapter{Conclusion and Future Work}
\label{cha:conc}

This study was undertaken to improve students' academic performance by using AI and semantic technologies. To achieve this goal, we came up with three objectives, along with three implementations, that is, to predict students' performance using grades from the previous semester, to model a course representation in a semantic way and compute the similarity, and to identify the sequence between two similar courses. For highlighting the research gaps in the Computer Science curriculum semantic analysis and contributing to the growing field of research, we conducted a systematic review regarding what semantic technologies are currently being used. A major finding of the study is that technologies used to measure similarity have limitations in terms of accuracy and ambiguity in the representation of concepts, courses, or curriculum. Our research fills this gap. Furthermore, this review also inspires us to think further regarding identifying the sequence between similar courses.

Regarding students' academic prediction, we conducted a dropout prediction experiment on a dataset from a Brazilian university. Three clean datasets were generated after pre-processing. Then, LSTM was performed in order to predict dropouts based on a SVM-based GA feature selection. Taken together, the results from the experiment illustrate that LSTM has strong adaptability in predicting dropout as there is breakthrough progress in terms of accuracy, which improves the best accuracy by 2.45\% over Manrique's work \citep{Ruben2019} on the ARQ dataset. Due to unbalanced datasets, we also observed the limitations of this study. The model has a bias in feature selection, abnormal accuracy decline during training, and multiple descents in loss, which emphasizes the importance of the balance of datasets.

With respect to course similarity measurement, we deployed BERT, which has the strong power of input embedding, to encode the sentence in the course description from the Australian National University. We then used cosine similarity to obtain the distance between courses. As a result, we found that COMP1110 has the closest average similarity to the rest of the 1000-level courses. In 2000-level courses, we identified COMP2100 has the closest relationship with the rest. In terms of courses at the 3000 and 4000 level, since specialization has formed, comparing them is pointless. The results also align with this viewpoint. As regards limitations, at this stage we cannot evaluate the result.  

The final step in this project is measuring the sequence between two similar courses. We employed the textRazor to extract entities from course description and then used SemRefD, which was proposed and evaluated by \citep{Ruben2019}, to measure the degree of prerequisite between two concepts. By deploying the model on COMP1100, COMP1110, and COMP2100 respectively, we established the relationship between these three courses. The results show that COMP1100 and COMP1110 have a strong prerequisite relationship with COMP2100, while the relationships between the two courses are inclined to be on the same level.

In terms of the future work, these technologies could potentially be used to analyse the curriculum of university programs, to aid student advisors, and to create recommendation systems that combine semantic and deep learning technologies. Furthermore, this study suggested that using GA+LSTM model could provide institutions with early detection for dealing with problems and retaining students. As for the experiment's future refinement, first, in dropout prediction, dynamic time steps can be introduced in LSTM instead of using fixed steps to make full use of the dataset. Moreover, a more balanced dataset could be used in the experiment. We will continue to investigate and develop appropriate models to improve the results in the future.


\backmatter

\cleardoublepage
\phantomsection
\addcontentsline{toc}{chapter}{Bibliography}
\bibliographystyle{anuthesis}
\bibliography{report}

\chapter{Appendix 1}
\label{cha:appendix1}

\textbf{\LARGE{Final Project Description}}
\\
\\
This project aims to improve students' performance by predicting students' dropouts and curriculum semantic analysis.

As for dropout prediction, this project uses SVM- based Genetic Algorithm to perform a feature selection; followed by using LSTM to predict the dropout. The details is regarding conducted  a  dropout  prediction experiment on a dataset from a Brazilian university. Three clean datasets were generated after pre-processing.  Then, LSTM was performed in order to predict dropouts based on a SVM-based GA feature selection.

With respect to course similarity measurement, we deployed BERT, which has the strong power of input embedding, to encode the sentence in the course description from the Australian National University.  We then used cosine similarity to obtain the distance between courses.

Regarding measuring  the  sequence  between  two  similar courses.  We employed the textRazor to extract entities from course description and then used SemRefD, which was proposed and evaluated by Manrique, to measure the degree of prerequisite between two concepts. By deploying the model on COMP1100, COMP1110, and COMP2100 respectively, we established the relationship between these three courses.

The outcomes of project are expected to use these technologies could potentially be used to analyse  the  curriculum  of  university  programs,  to  aid  student  advisors,  and  to  create recommendation  systems  that  combine  semantic  and  deep  learning  technologies.Furthermore, this study suggested that using GA+LSTM model could provide institutions with early detection for dealing with problems and retaining students. 

\chapter{Appendix 2}
\label{cha:appendix2}
\textbf{\LARGE{Contract}}

\begin{figure}
  \centering
  \includegraphics[width=\textwidth]{{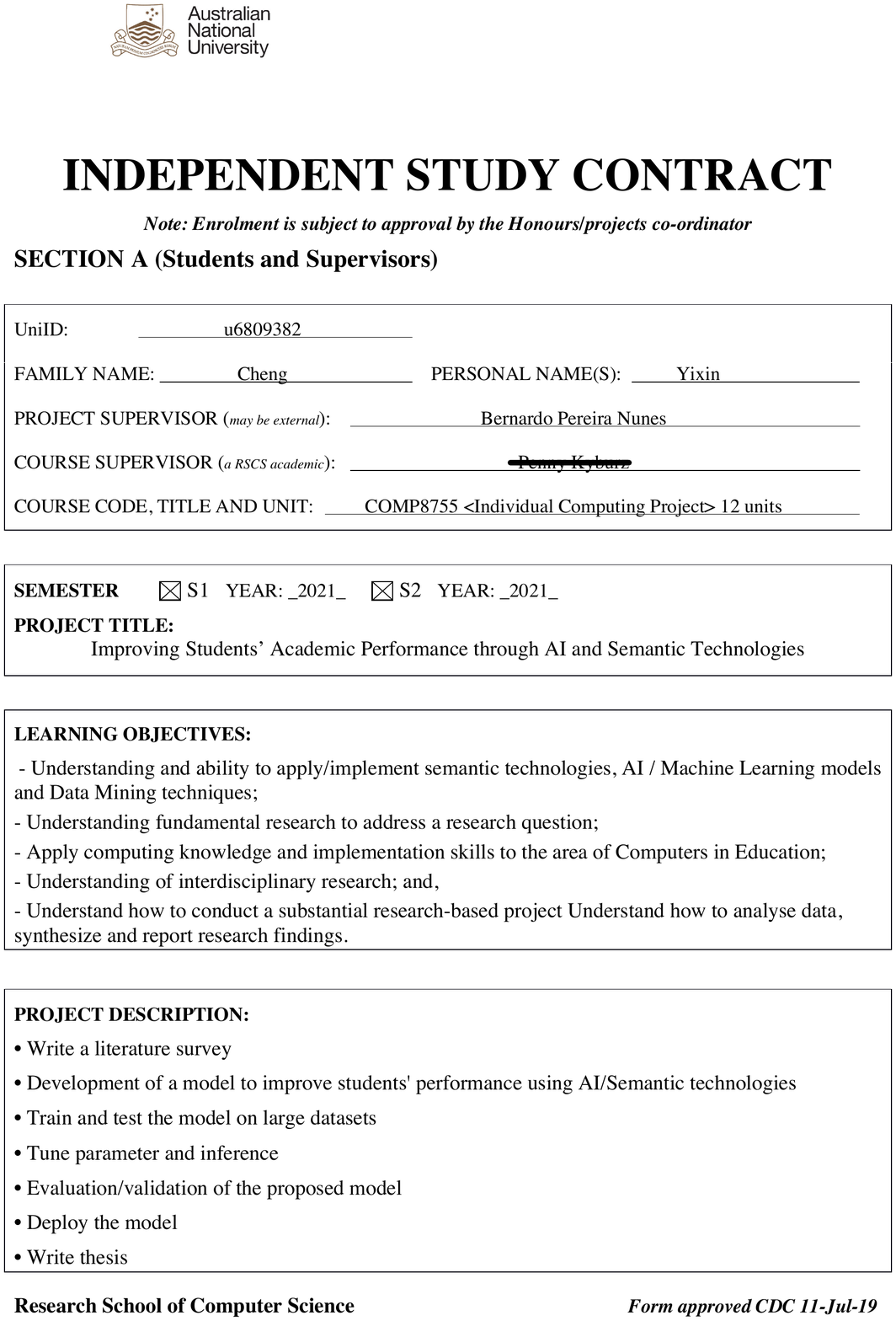}}
  \caption{Contract page 1}
  \label{fig:semrefd}
\end{figure}

\begin{figure}
  \centering
  \includegraphics[width=\textwidth]{{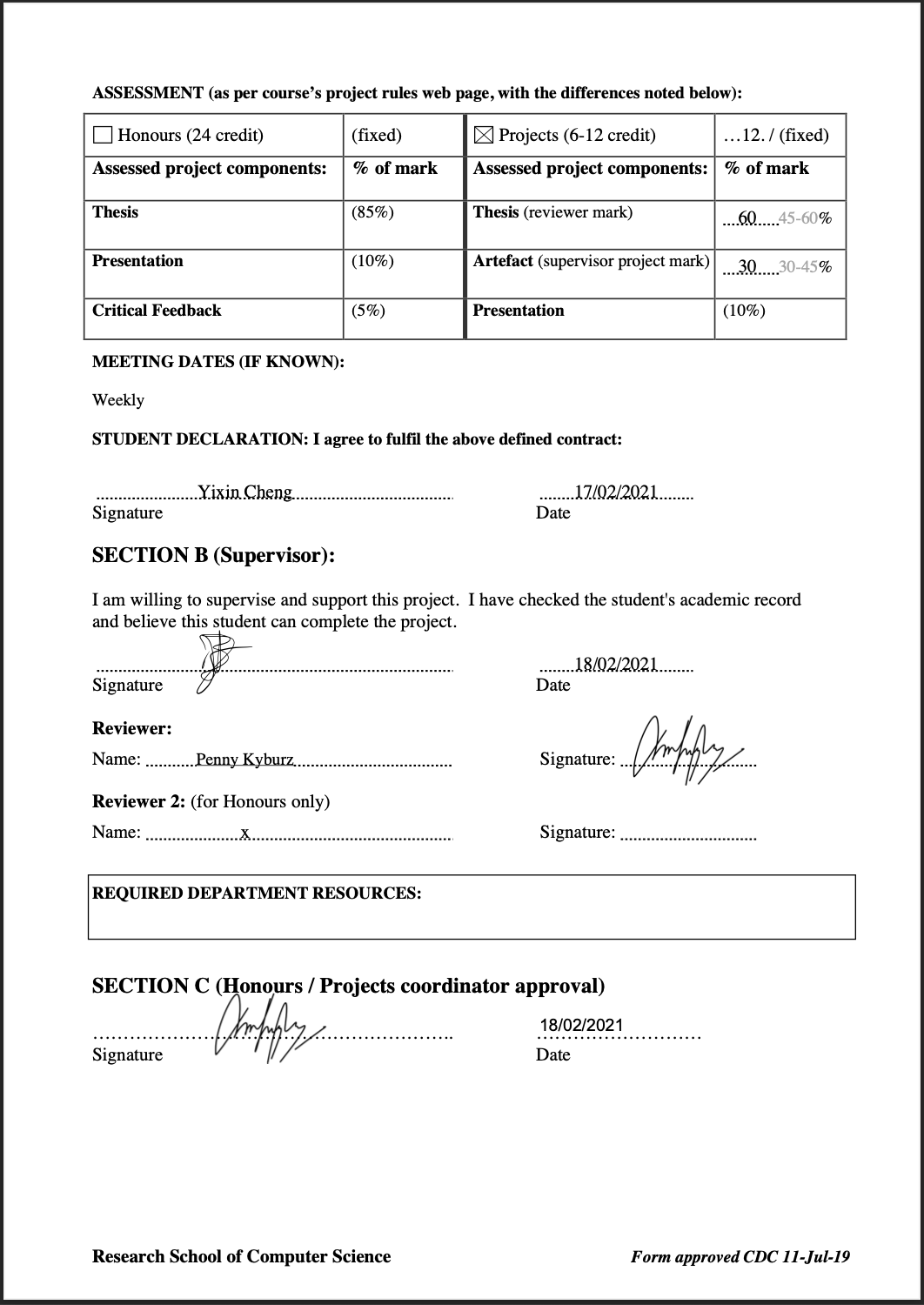}}
  \caption{Contract page 2}
  \label{fig:semrefd}
\end{figure}
\chapter{Appendix 3}
\label{cha:appendix3}
\textbf{\LARGE{Description of Software}}
\\
\\
\textbf{\large{Dropout prediction}}
\\
\\
preprocessing.py implements the pre-processing. subset\_seperator(self) is separating raw dataset by degree; organise(self) is for data cleaning, data normalization and Data validation. impute\_missing\_value(self) is for data imputation.
\\
\\
feature\_selector.py implements the feature selection by using GA+SVM. get\_fitness (self,pop,path) is for getting fitness of each individual by SVM; select(self,pop, fitness), crossover(self,parent, pop), mutate(self,child), and evolution(self) are the steps of GA.
\\
\\
xx\_dataloader.py, xx=ADM,ARQ, or CSI implements the training and test by LSTM.
\\
\\
Directly Run xx\_dataloader.py, xx=ADM,ARQ, or CSI to get the result of preprocessing, feature selection and training and test.
\\
\\
\textbf{\large{Similarity Measurement}}
\\
\\
Installment of BERT as Service and runing
The installment of BERT used in this project as the link above or following.
\\
\\
Install Bert as a Service by pip install -U bert-serving-client(instruction is available on https://github.com/hanxiao/bert-as-service)
\\
\\
Run the Bert as a Service pointing to the unzipped downloaded model using the following command: bert-serving-start -model\_dir /your\_directory/wwm\_uncased\_L-24\_H-1024\_A-16 -num\_worker=4 -port XXXX -max\_seq\_len NONE
\\
\\
run python similarity.py to get the result of two sentences in two courses comparison. The result will be in /result/similarity\_measurement/full\_similarity
\\
\\
run python text\_similarity.py to get the result of two courses comparison, which will be stored in /result/similarity\_measurement/full\_similarity/similarity\_full.csv
\\
\\
run python diagram.py to visualize the result.
\\
\\
\textbf{\large{Prerequisite Identification}}
\\
\\
Install the external libraries by using conda or pip command
RefDSimple.py is for using RefD to retrieve and get potential candidates from DBpedia https://www.dbpedia.org/ which is a main knowledge graph in Semantic Web.
\\
\\
entity\_extractor.py is using TextRazor (https://www.textrazor.com/) to segment and extract entities from text.
\\
\\
config.cfg is the configuration file, you may need to change the proxy before using it.
\\
\\
run entity\_extractor.py to get the result which will be in /result/prerequisite\_identification
\\
\\
\textbf{\large{Systematic Review-Paper Crawling}}
\\
\\
Run paper\_crawler.py to acquire the paper from SPringer to local file named "datasetbyabstract", which includes title, abstract, and so on.
\chapter{Appendix 4}
\label{cha:appendix4}
\textbf{\LARGE{README file}}

\begin{figure}
  \centering
  \includegraphics[width=\textwidth]{{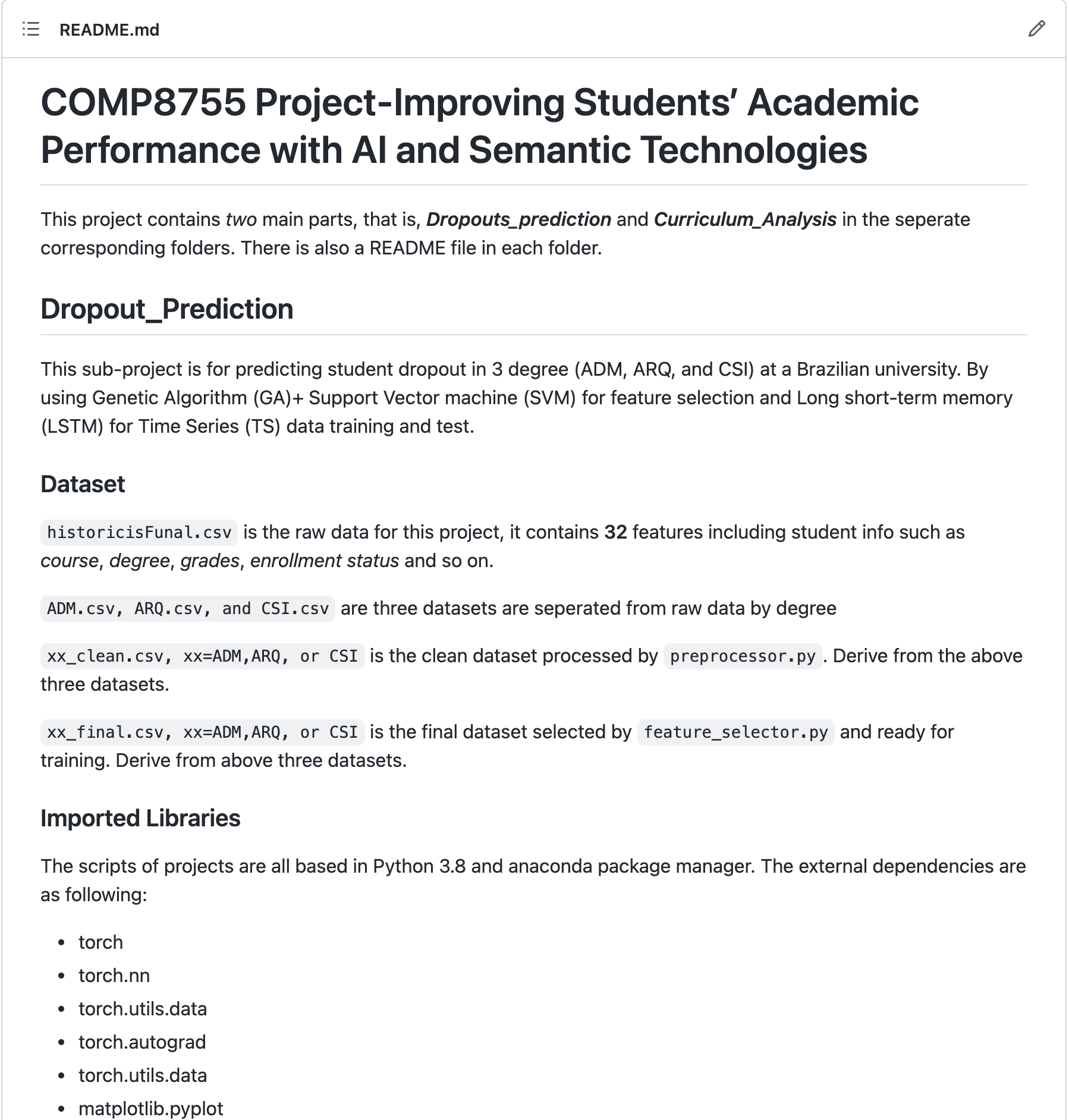}}
  \caption{README page 1}
  \label{fig:1}
  
\end{figure}
\begin{figure}
  \centering
  \includegraphics[width=\textwidth]{{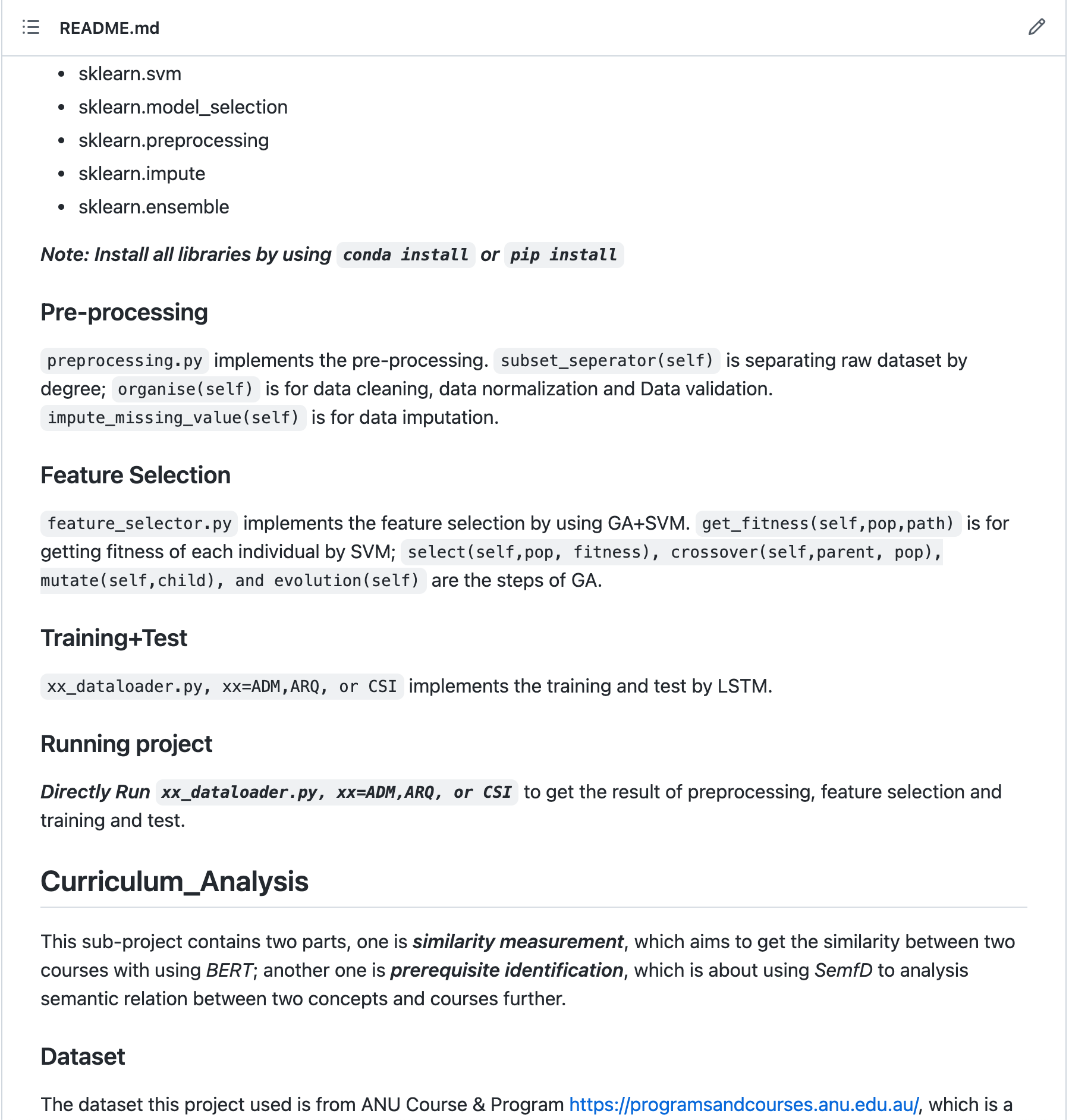}}
  \caption{README page 2}
  \label{fig:2}
  
\end{figure}
\begin{figure}
  \centering
  \includegraphics[width=\textwidth]{{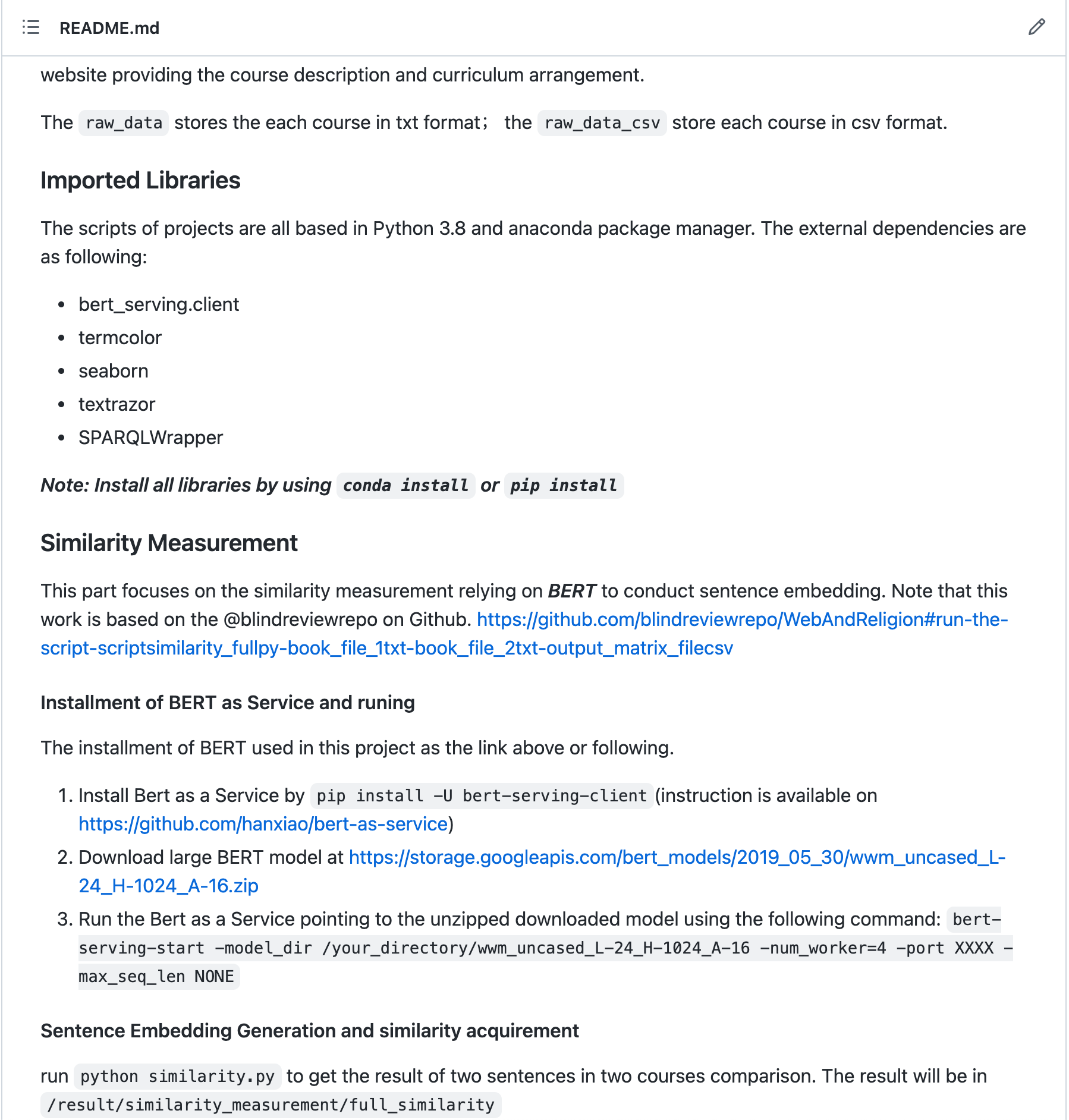}}
  \caption{README page 3}
  \label{fig:3}
\end{figure}
\begin{figure}
  \centering
  \includegraphics[width=\textwidth]{{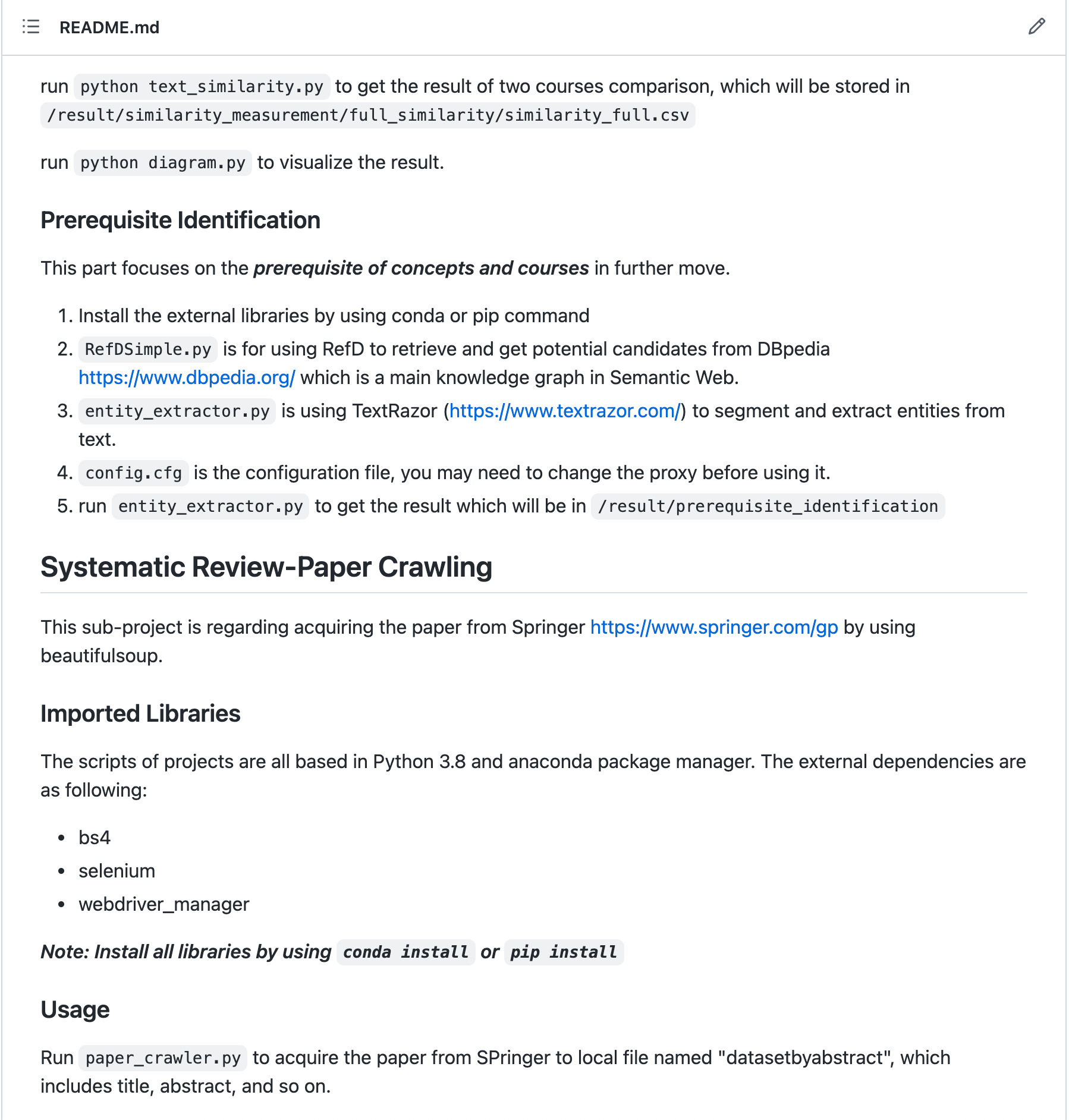}}
  \caption{README page 4}
  \label{fig:4}
\end{figure}

\printindex

\end{document}